\documentclass[conference]{IEEEtran}
\IEEEoverridecommandlockouts % Only needed for the first footnote. Comment out if not needed.

\usepackage{cite}
\usepackage{amsmath,amssymb,amsfonts}
\usepackage[us,12hr]{datetime}
\usepackage{graphicx}
\usepackage{textcomp}
\usepackage{xcolor}
\usepackage{xparse}
\usepackage{comment}
\usepackage{float}
\usepackage{colortbl}
\usepackage{booktabs}
\usepackage{nameref}
\usepackage{hyperref}
\usepackage[strings]{underscore}
\usepackage{tabularx}
\usepackage{multirow}
\usepackage{layouts}
\usepackage{wasysym}

\usepackage{url}

\usepackage{etoolbox}
\apptocmd{\sloppy}{\hbadness 10000\relax}{}{}

\usepackage{balance}

\begin{document}

%%%%%%%%%%%%%%%%%%%%%%%%%%%%%%%%%%%%%%%%%%%%%%%%%%%%%%%%%%%%%%%%%%%%%%%%%%%%%%%%%%%%%%%%%%%%%%%%%%%%
\title{Best Practices for IoT Security:\\What Does That Even Mean?
    \thanks{Version: 24 April 2020. Preliminary results.}
}

\author{
    \IEEEauthorblockN{Christopher Bellman $~~~~$ Paul C. van Oorschot}
	\IEEEauthorblockA{\textit{School of Computer Science} \\
		\textit{Carleton University}, Ottawa, Canada  
	}
}

\maketitle

%These next two lines add page numbers.
\thispagestyle{plain}
\pagestyle{plain}

%%%%%%%%%%%%%%%%%%%%%%%%%%%%%%%%%%%%%%%%%%%%%%%%%%%%%%%%%%%%%%%%%%%%%%%%%%%%%%%%%%%%%%%%%%%%%%%%%%%%
\begin{abstract}    
    Best practices for Internet of Things (IoT) security have recently attracted considerable attention worldwide from industry and governments, while academic research has highlighted the failure of many IoT product manufacturers to follow accepted practices. We explore not the failure to follow best practices, but rather a surprising lack of understanding, and void in the literature, on what (generically) ``best practice'' means, independent of meaningfully identifying specific individual practices. Confusion is evident from guidelines that conflate desired outcomes with security practices to achieve those outcomes. How do best practices, good practices, and standard practices differ? Or guidelines, recommendations, and requirements? Can something be a best practice if it is not \textit{actionable}? We consider categories of best practices, and how they apply over the lifecycle of IoT devices. For concreteness in our discussion, we analyze and categorize a set of 1014 IoT security best practices, recommendations, and guidelines from industrial, government, and academic sources. As one example result, we find that about 70\% of these practices or guidelines relate to early IoT device lifecycle stages, highlighting the critical position of manufacturers in addressing the security issues in question. We hope that our work provides a basis for the community to build on in order to better understand best practices, identify and reach consensus on specific practices, and then find ways to motivate relevant stakeholders to follow them.
\end{abstract}

\begin{IEEEkeywords}
Internet of Things, IoT, Best Practices
\end{IEEEkeywords}

%%%%%%%%%%%%%%%%%%%%%%%%%%%%%%%%%%%%%%%%%%%%%%%%%%%%%%%%%%%%%%%%%%%%%%%%%%%%%%%%%%%%%%%%%%%%%%%%%%%%
\section{Introduction}
When we talk of ``best practices'', we rely on each other's intuitive understanding of what a best practice is. Unfortunately, closer inspection suggests that we lack both a good understanding of this term, and an unambiguous explicit definition. In this paper,  we provide what we believe is the first in-depth study towards understanding what is meant by ``security best practice''. While for concreteness herein our main scope is security and privacy best practices for consumer-focused Internet of Things (IoT), we believe that much of our work may be of independent interest beyond IoT or security.

IoT is commonly described as adding network connectivity to typically non-networked items or ``things'' \cite{Wortmann2015}. It surrounds us with a variety of network-connected devices such as smart light bulbs, door locks, and less obvious objects like fridges, toasters, traffic lights, or sensors and controllers built into critical infrastructure systems. The importance of IoT in marketing and sales has resulted in the production of a wide variety of devices with arguably ``unneccessary'' functionality (e.g., internet-connected toasters and stuffed animal toys). These devices, while perhaps convenient and highly functional for users, have acquired a reputation \cite{Alrawi2019} of poor security and misconfiguration, leading to huge numbers of network-accessible devices being vulnerable. As IoT devices may be more isolated or resource-constrained (e.g., battery power, processors, memory), or lacking in software update support, their security issues are often hard to address. The cyberphysical nature of IoT---interfacing to physical world objects---results in threats to our physical world, as well as to networks and other internet hosts \cite{Kolias2017}. This has resulted in attention to identifying best practices for IoT security.

The concept of ``best practices'' appears rather nebulous and informal. Academic work on this is scant and lacks concrete definitions, relying instead on apparently reasonable common sense. For example, in considering Cloud Security Providers (CSPs), Huang et al. \cite{huang2015} refer to: ``security mechanisms that have been implemented across a large portion of the CSP industry [are thus] considered standardized into a `best-practice'.'' Here best practice appears to mean \textit{widely implemented}. In their evaluation of home-based IoT devices, Alrawi et al. \cite{Alrawi2019} note numerous violations of security design principles, and assert ``Best practices and guidelines for the IoT components are readily available'', but cite none among 108 references, and offer no definition.  In a recent national news article \cite{CBC2020} on banks disclaiming liability for customer losses from e-transfer fraud, and one-sided online banking agreements, a defensive bank representative is quoted: ``We regularly review our policies and procedures to ensure they align with best practices.'' This quote appears to be not about security best practices, but rather legal best practices in the sense of \textit{our agreements are no worse than our competitors}. Large collections of documents from industrial, government, and academic sources also suggest conflation of the term ``best practice'' with other common terms such as ``recommendation'' and ``guideline'' \cite{DCMS1}. These examples collectively suggest confusion, ambiguity, and misuse resulting from the lack of a common understanding or precise definition of the term ``best practice''.

%Contribution
We first explore current use of terms related to best practices, and how their meanings differ qualitatively. We organize these descriptive terms into three categories, giving a visual model as aid. We distinguish and define (actionable) security \textit{practices} distinct from \textit{desired security outcomes}, which we further split into type-S (specific outcomes suggesting practices) and type-V (vague outcomes). To get a sense of the stakeholders potentially involved in best practices, and when, we use a lifecycle model for IoT security devices. We use this, and thirteen ``Guidelines'' from the UK's \textit{Code of Practice for Consumer IoT Security} \cite{DCMS2}, as a basis for our own analysis of over 1000 items from a companion UK government document \cite{DCMS1} which maps these items onto the thirteen guidelines. The rich collection of over 1000 items is from industry and government sources offering IoT security and privacy best practices, best available methods, security guidelines, requirements, recommendations, and standards. Our analysis finds that for consumer IoT security, the majority of practices relate to early phases of an IoT device's lifecycle. As a result, poor security posture developed early in the lifecycle accrues what may be called a ``security debt'' (cf. \textit{tech debt}---easy but low quality technical choices made during development incur later costs \cite{Kruchten2012}). We also find  that an overwhelming majority of recommendations (91\%) are not actual practices but rather  desired outcomes.

This suggests that many listed recommend practices are too vague for the relevant stakeholders to implement successfully, which then negatively impacts stakeholders at later stages in a lifecycle. An observation that follows easily is that all stakeholders seeking advice on best practices would benefit if guideline lists were reworked to also suggest examples of suitable practices (rather than desired outcomes), or if lists of desired outcomes were supplemented by examples of actionable practices known to reliably achieve the outcome. 

%%%%%%%%%%%%%%%%%%%%%%%%%%%%%%%%%%%%%%%%%%%%%%%%%%%%%%%%%%%%%%%%%%%%%%%%%%%%%%%%%%%%%%%%%%%%%%%%%%%%
\section{Confusion About what ``Best Practice'' Means}
\label{sec:bestpractices}

Here we offer a definition for ``best practice'', and consider the concepts of outcomes, actions, and ``actionable'' practices. We also discuss related terms commonly appearing in literature. Through this, we aim for a more precise working vocabulary in discussing best practices, and to disambiguate the wide variety of qualifying terms into meaningful categories.

The definition of ``best practice'' seems to be taken somewhat for granted, as very few documents that use it make an effort to define it. Of note, even RFC 1818/BCP 1 \cite{rfc1818}, the first of the IETF RFCs specifying what a Best Current Practice document is, fails to define ``best practice''. This suggests the term (and concept of) best practice is used casually, versus scientifically---presumably everyone understands what it means well enough to not require a specific definition. This, however, leads to ambiguity, where certain uses of ``best practice'' have different meanings and connotations, while elsewhere different phrases may imply the same concept. 

While focused more on human aspects of best practices, of specific note is King's discussion \cite{King2000} of security best practices, where it is defined as ``practices that have proven effective when used by one or more organizations and which, therefore, promise to be effective if adapted by other organizations''. King's discussion of the definition covers a number of important concepts, including that effectiveness is based on evidence of multiple instances (i.e., a consensus), the practice must be applicable to real situations, not theoretical; and that it may exist among a set of practices of equal quality to perform a security process \cite{King2000}.

In an effort to reduce ambiguity in terminology, we suggest the following working definition: \textit{For a given desired outcome, a ``best practice'' is a means intended to achieve that outcome, and that is considered to be at least as ``good'' as the best of other broadly-considered means to achieve that same outcome}. It is something that can be practiced (an action), not something that is desired to be achieved (an outcome). 

It appears infeasible to specify an \textit{objective} best practice, i.e., one that is best possible for all environments and applications and target users. A determination of an objective best practice would require knowledge of not only every practice and some way to measure their quality for comparison with one another, but also a one-size-fits-all practice that addressed all circumstances. In lieu of an objective best practice, we consider a best practice to be born of a consensus (if reachable), e.g., of the experts in a given field. Some degree of consensus or common use within a group may suffice for a practice to considered a best practice if it is seen as better than other means to achieve the same goal (from our definition). As such, there may be different political pressures into designating a practice as a best practice. A stakeholder in some practice (e.g., manufacturer of a product to use in a practice) may try to build consensus that their practice is the best for their own gain (economic or otherwise).

%Can there be multiple best practices?
Can there be more than one best practice for a desired outcome? Our working definition suggests that a practice be \textit{at least as good} as others to achieve the same outcome. While one view of ``best'' might imply being above all known others, another is that ``best'' is a category that may have more than one member. We argue that it is reasonable to allow (by definition) that there are multiple best practices.

%Should a best practice indicate a goal?
The description of a best practice often fails to indicate an explicit desired outcome. We view this as a flaw; our working definition presupposes a desired outcome, and we suggest that a best practice for describing best practices is to explicitly identify the desired outcome. This may not be essential to implement the practice, but provides supporting information for an implementer to understand the objectives of their efforts.

%Are they always international/global; can they differ by country/culture?
Best practices, if defined by the groups that use them, can then differ across country or culture (and in some cases perhaps also across environments, applications and budgets). Even if we consider that an objective best practice exists if it can be formally verified as achieving an outcome in a superior way (often difficult given subjective metrics), groups unaware of such a practice would maintain their own practices that are, from their perspective, still the best. 

%Three angles: Legal, technical, social.
We can also consider the concept of a best practice from three angles: legal, technical, and social. From a legal perspective, following a best practice or standard may be used as an argument to escape or limit liability, as in ``following the crowd'' or consensus as surely being reasonable. An example is financial institutions citing ``industry best practices'' to disclaim liability, per our introduction \cite{CBC2020}. Technically, a best practice is the best way known to technical experts or researchers, for achieving an outcome (as agreed by some form of consensus). Socially, ``best practice'' often implies the most common (if not necessarily best) way to do something. At one level, one might argue that each of these are similar, but at a deeper level, their semantic meanings are quite different uses of the same term. 

    \subsection{Outcomes vs. Actions}
    \label{sec:background/actionvsoutcome}
    \label{sec:background/actionable}
    An \textit{outcome} is the end goal that a stakeholder desires to reach. An \textit{action} is the technical means by which to reach an outcome. For example, an outcome may be to ``store user passwords securely'', and an action to achieve this outcome may be ``salt and hash user passwords using current NIST-approved algorithms''. In practice, outcomes or goals that are vague or broad may not give stakeholders a clear idea of any concrete set of actions that would achieve the goal. A desired outcome of ``strong security'', for example, is nebulous, and cannot be mapped to specific actions to achieve the goal. It is folly to attempt to determine actions that achieve an intangible goal. In contrast, defining tightly-scoped outcomes or specifying an objective to withstand specific attacks, allows for successful mapping to corresponding actions.
    
    A best practice, by our working definition, must be actionable in order to qualify as a practice. An outcome can not be a best practice via this same definition, as it does not specify a means to an end. Once an action is defined, the practice may be viewed as \textit{actionable}, meaning that it can be acted upon by an implementer. We define an ``actionable'' practice as \textit{a practice that involves a known sequence of steps, and it is known how to do them}. An ``implementer'' is a stakeholder (manufacturer, user, or otherwise) who would be responsible for or be in the best position to implement a practice.  Implementers should be able to understand what is required to carry out a practice. This means wording and outcomes must be understood from both a (semantic) language perspective and, importantly, a technical perspective, and the practices should involve only available existing techniques. Requiring techniques that are experimental or unproven introduces ambiguity in how to carry out a practice and results in inconsistent implementations.
    
    Some outcomes more clearly suggest actions than others (e.g., ``no default passwords'' vs. ``validate input data'' \cite{DCMS2}). For this reason we categorize outcomes into two broad types:

%%%%%%%%%%%%%%%%%%%%%%%%%%%%%%%%%%%%%%%%%%%%%%%%%%%%%%%%%%%%%%%%%%%%%
%Vars:
\newcommand{\LenClass}{2.5cm}
\newcommand{\LenTerm}{3.0cm}

\newlength\LenMax
\setlength\LenMax\textwidth
\addtolength\LenMax{-\LenClass}
\addtolength\LenMax{-\LenTerm}
\addtolength\LenMax{-1cm}

%%%%%%%%%%%%%%%%%%%%%%%%%%%%%%%%%%%%%%%%%%%%%%%%%%%%%%%%%%%%%%%%%%%%%

\begin{table*}[t]
    \centering
    \caption{A classification of commonly-used general qualifying terms for practices and their suggested usage in literature.}
    \label{tab:qualifyingterms}
    \begin{tabular}{@{}p{\LenClass}p{\LenTerm}p{\LenMax}@{}}
        \toprule
        Category & 
        Terms & 
        Suggested Usage \\ \midrule
        %%%%%%%%%%%%%%%%%%%%%%%%%%%%%%%%%%%%%%%%%%%%%%%%%%%%%%%%%%%%%%%%%%%%%%%%%%%%%
        \"{U}ber Practices& ``above-and-beyond'' & \multirow{2}{\LenMax}{For practices that provide superior outcomes, though not widely adopted. These terms tend to imply top-tier quality, albeit sometimes at very high cost or complexity.}\\
        & ``gold standard'' & \\ 
        & ``state-of-the-art'' & \\ \addlinespace[0.05cm] \midrule
        %%%%%%%%%%%%%%%%%%%%%%%%%%%%%%%%%%%%%%%%%%%%%%%%%%%%%%%%%%%%%%%%%%%%%%%%%%%%%
        %Class 1 & ``best practice'' &\\
        Best Practices& ``best practice'' & \multirow{3}{\LenMax}{Used when describing the practices widely-considered to be the highest level of quality. These are \textit{widely-considered} (plus commonly adopted, typically) whereas for above-and-beyond, wide consideration or broad use is not a requirement.}\\
        & ``best current practice'' & \\
        &  & \\ \addlinespace[0.05cm] \midrule
        %%%%%%%%%%%%%%%%%%%%%%%%%%%%%%%%%%%%%%%%%%%%%%%%%%%%%%%%%%%%%%%%%%%%%%%%%%%%%
        Good Practices& ``good practice'' & \multirow{2}{\LenMax}{For practices that are beneficial to implement and improve quality (versus not implementing it), without implying that better practices do not exist.}\\
        & ``suggested practice'' &\\
        & ``recommended practice'' &\\
        & ``acceptable practice'' & \\\addlinespace[0.05cm] \midrule\midrule
        %%%%%%%%%%%%%%%%%%%%%%%%%%%%%%%%%%%%%%%%%%%%%%%%%%%%%%%%%%%%%%%%%%%%%%%%%%%%%
		Common Practices& ``common practice'' & \multirow{2}{\LenMax}{For practices that do not necessarily imply the quality of an associated practice, but instead suggest wide use.}\\
		& ``standard practice'' & \\
		& ``accepted practice'' & \\\addlinespace[0.05cm] \midrule
		%%%%%%%%%%%%%%%%%%%%%%%%%%%%%%%%%%%%%%%%%%%%%%%%%%%%%%%%%%%%%%%%%%%%%%%%%%%%%
        Requirements& ``requirement'' & \multirow{3}{\LenMax}{For practices that are, in some way, enforced by an entity such that there implies a negative consequence should the practice not be followed. Alternatively, these may be de facto practices or functionality, informally recognized by experts as essential.}\\
        & ``minimum requirement'' & \\
        & ``mandatory practice'' & \\
        & ``baseline practice'' & \\ 
        & ``code of practice'' & \\
        & ``regulation'' & \\ \addlinespace[0.05cm]\midrule\midrule
		%%%%%%%%%%%%%%%%%%%%%%%%%%%%%%%%%%%%%%%%%%%%%%%%%%%%%%%%%%%%%%%%%%%%%%%%%%%%%
		Formal Recognition& ``formal standard'' & \multirow{1}{\LenMax}{For practices that are endorsed in some official capacity by an organization or individual.}\\
		& ``recommendation'' & \\
		& ``guideline'' & \\ 
		& ``guidance'' & \\\addlinespace[0.05cm]
		%%%%%%%%%%%%%%%%%%%%%%%%%%%%%%%%%%%%%%%%%%%%%%%%%%%%%%%%%%%%%%%%%%%%%%%%%%%%%
        \bottomrule
    \end{tabular}
\end{table*}

    \begin{itemize}
        \item Type-V outcome (Vague)---the outcome does not strongly imply actions to take to achieve it 
        \item Type-S outcome (Specific)---the outcome implies specific actions to take to achieve it    
    \end{itemize}
    
    Type-S outcomes tend to involve a specific goal, whereas type-V outcomes tend to involve broad or vague goals. Type-S outcomes are more useful in practice, with less room for (mis)interpretation of what is required to achieve the outcome. Upon deeper inspection, a number of sub-categories would emerge. This depth is not further explored herein.
    
    It follows that a recommendation specifying an outcome, but the path to which is an open research problem, cannot (and should not) be considered an actionable practice. It is important for the security (in our case) community---whether by academic, industrial, or government efforts---to identify and agree on actionable practices with concrete desired outcomes for implementers. These can be adopted or standardized across applications and implementations, thereby simplifying implementation and establishing  confidence in the authorities that recommend and endorse best practices.
    
    %Discuss external to the definition of actionable the idea that resources available is highly related.
    We separate the two concepts of a practice being actionable, and an implementer having the means by which to put said practice into place. Implementers must have the resources (technical, financial, personnel) available before a practice can be implemented, but availability of resources (or lack thereof) does not affect the generic actionability of a practice. (Note: though a practice is actionable in general, that does not guarantee that a given party themself has the resources to adopt the practice.)  A practice that has a significant cost may be ruled out as a best practice by a recommending group, governing body, or peer community.  
    
    \subsection{Commonly-Used Qualifying Terms}
    A number of ``qualifying terms'' describing the quality of some practice (e.g., ``common'', ``good'', ``best'') are often used without definition or seemingly interchangeably within literature. This lack of clarity calls for a careful examination of their meaning and usage. Being widely used in literature might suggest that readers know (and are in universal agreement on) what authors mean when they use one of these terms, but this appears false---and the root cause is, we suggest, the absence of explicit definitions. In an effort to both highlight existing, and reduce ongoing ambiguity, we categorize a number of commonly-used qualifying terms into classes and suggest where/when each term should be used. To this end, Table~\ref{tab:qualifyingterms} outlines commonly-used qualifying terms, their associated usage categories, and a brief description of their usage. This classification describes each \textit{term} and what categories of practices are associated with them. Regardless of category that a qualifying term may fall into, there remains the requirement that any (true) ``practice'' as defined by these terms should  be actionable, for reasons as discussed earlier.
    
    \subsection{Practice Quality}
    Quality-based terms provide a natural basis on which to differentiate practices, e.g., ``good'' or ``best''. Here we delineate three categories ranging from the highest possible quality of a practice, to those that still improve security posture but are not widely considered the best.
    
        \paragraph*{``\"{U}ber Practices''}
        The term ``\"{U}ber Practices'' suggests practices in some way superior to (other) ``best'' practices, or somehow beyond what would be considered already high quality. Similarly, ``state-of-the-art'' implies something at an absolute peak of technical quality, but perhaps not yet widely adopted. Consider as a practical example: in luxury cars, a heated steering wheel. While more comfortable on a winter day, best practice would likely be to ensure correct function and adequate steering grip to reduce likelihood of accidents. A heating function is ``above and beyond'' that.
        
        \paragraph*{``Best Practices''}
        ``Best practice'' terms describe practices widely considered to be both the best available and relatively widely adopted. A key point is wide consideration among a group as being a best practice. While practices may exist that are technically better (\"{U}ber practices), best practices are widely accepted to be high in quality.
        
        \paragraph*{``Good Practices''}
        ``Good practices'' improve the security posture, but are not necessarily the best security practices available. They generally are not lauded for high quality per se. A ``good'' practice often either does not have wide acceptance as being the best, or is perhaps not widely practiced or not considered essential despite being easy and beneficial (e.g., turning the wheels to the curb when parking on a hill). Further consideration for context may prove useful for understanding good-practice usage. For example, securing a low-value free online newspaper account may not require a best practice (as per the above definition), but a good practice that adequately fulfils its task. In other words, \cite{King2000}: ``sometimes the good is good enough''.
        
        Conceptually, we categorize ``good'', ``best'', and ``\"{u}ber'' along one continuum or dimension of quality, ranging from lower to higher.
    
        We note that terms used to describe practices of low quality (i.e., below good) receive less attention in literature as documents promoting security advice focus more on good than bad practices. Our definition of a good practice (the lowest quality we formally categorize) implies that anything lower does not improve security posture.
    
    \subsection{Common Practices and Requirements.}
    Common practices and requirements share with quality terms (above) their trait of having ``practice'' commonly within their terms (e.g., ``accepted practice'', ``mandatory practice''). However, in contrast, they suggest frequency or obligation of use.
    
        \paragraph*{Common Practices}
        ``Common practice'' qualifying terms do not imply any quality of associated practices, but rather reflect broad usage. For example, it may (unfortunately) be common to store passwords in plaintext within a database (thus being a common practice), but that is not best practice (or even a good practice).
        
        \paragraph*{Requirements} A requirement differs from quality-based terms (\"{u}ber, best, good) as it does not necessarily imply the quality of a given practice, but rather that it is mandated by some governing body or general expectation. They may be considered ``enough'' for some purposes (e.g., ``enough to not be sued'' or ``enough to pass inspection/regulation''). Practices across a range of qualities may be requirements depending on the implied governing body or motivation. Requirements tend to sit somewhere between common practices and formal recognition. On one hand, their terminology may contain ``practice'' (e.g., mandatory practice, baseline practice) without implying a measure of quality, but on the other hand, they are required by some governing body, suggesting formal recognition.
        
        Because the ``requirements'' and ``common practice'' categories do not necessarily define quality, we assign them to an orthogonal ``commonality'' dimension, with common practice below requirement. Correlated to commonality is the \textit{maturity} of a practice. A practice that has proven to be of a high quality is more likely to become a requirement or part of a standard. A requirement is often just a binary ``yes'' or ``no'', but may also be non-binary, e.g., as an ordering within a large set of related requirements. Practices may be requirements or common practices, and a common practice or requirement may be a good, best, or \"{u}ber practice. Ideally, a best practice will become a common practice, and both become required, and endorsed (below).
    
    \subsection{Formal Recognition (Endorsement)}
    While the above discusses quality, commonality, and maturity, other terms that arise surrounding best practices have more to do with informal or formal endorsements of a practice. This may trigger greater adoption (and is often the goal of an endorsement).  
    
        \paragraph*{Formal Standard}
        We use ``formal standard'' to mean a formalized (i.e., documented, agreed-on by some formal body) set of requirements. This implies not necessarily quality, but an established set of practices formalized for adoption by some group or governing body. The purpose of a standard may be interoperability, rather than quality per se---e.g., standards for the width of rail tracks, or widths of screws or pipes. In this context, ``Formal Standard'' differs from ``standard practice'' (i.e., common practice), which falls within the earlier discussion related to frequently practiced, e.g., it is \textit{standard} to eat at 12noon. Standards are typically sufficiently detailed such that conformance or compliance is easily judged by, e.g., an auditor, or interoperability tests.
    
        \paragraph*{Recommendation}
        A ``recommendation'' is an endorsement of, e.g., a practice by an entity (individual or organization)  as their suggested way to do something. Recommendations (depending on the recommending entity) may be subject to bias, and do not necessarily reflect expertise or universal consensus. For example, a government may recommend washing your hands with soap under warm water for 20 seconds \cite{CANHandWashing}---this may or may not be a best practice, but it is recommended by an organization. Recommendations are often less strict than requirements, and typically not as well documented as formal standards, but these fall along a common dimension also.  Some recommendations, by virtue of the body they arise from, are understood to be, in essence, requirements.  Recommendations commonly suggest following a standard \cite{DCMS1}. Just as we suggest that a best practice have an explicitly stated goal, a recommendation should, in our view, ideally be accompanied by an explanation of the intended outcome, to help implementers understand the motivation.  
        
        \paragraph*{Guidance/Guideline}
        A guideline or guidance is often given by an entity to promote a suggested way \textit{to achieve a goal}. This may be used as a synonym for a recommendation (above), positioned as guiding towards to a goal---or perhaps offered as help, versus imposing rules. We typically guide someone to something, suggesting a target, goal or desired outcome, ideally are explicitly stated if not obvious.
        
        The ``formal recognition'' category notes qualifying terms that describe endorsed practices. These too can be ordered along a continuum, with a \textit{formal standard} being on the upper end, most highly endorsed item; to \textit{guidance}, which may logically be considered as the lowest form of endorsement. Much like requirements and common practices, formal recognition/endorsements do not necessarily imply quality; however, the endorsing organization may have a recognized governing or expert status. This continuum related to endorsement can be placed on a third axis.
    
    \subsection{Conceptual Model of Terms}
\begin{figure}[t!]
        \centering
        \includegraphics[width=0.40\textwidth]{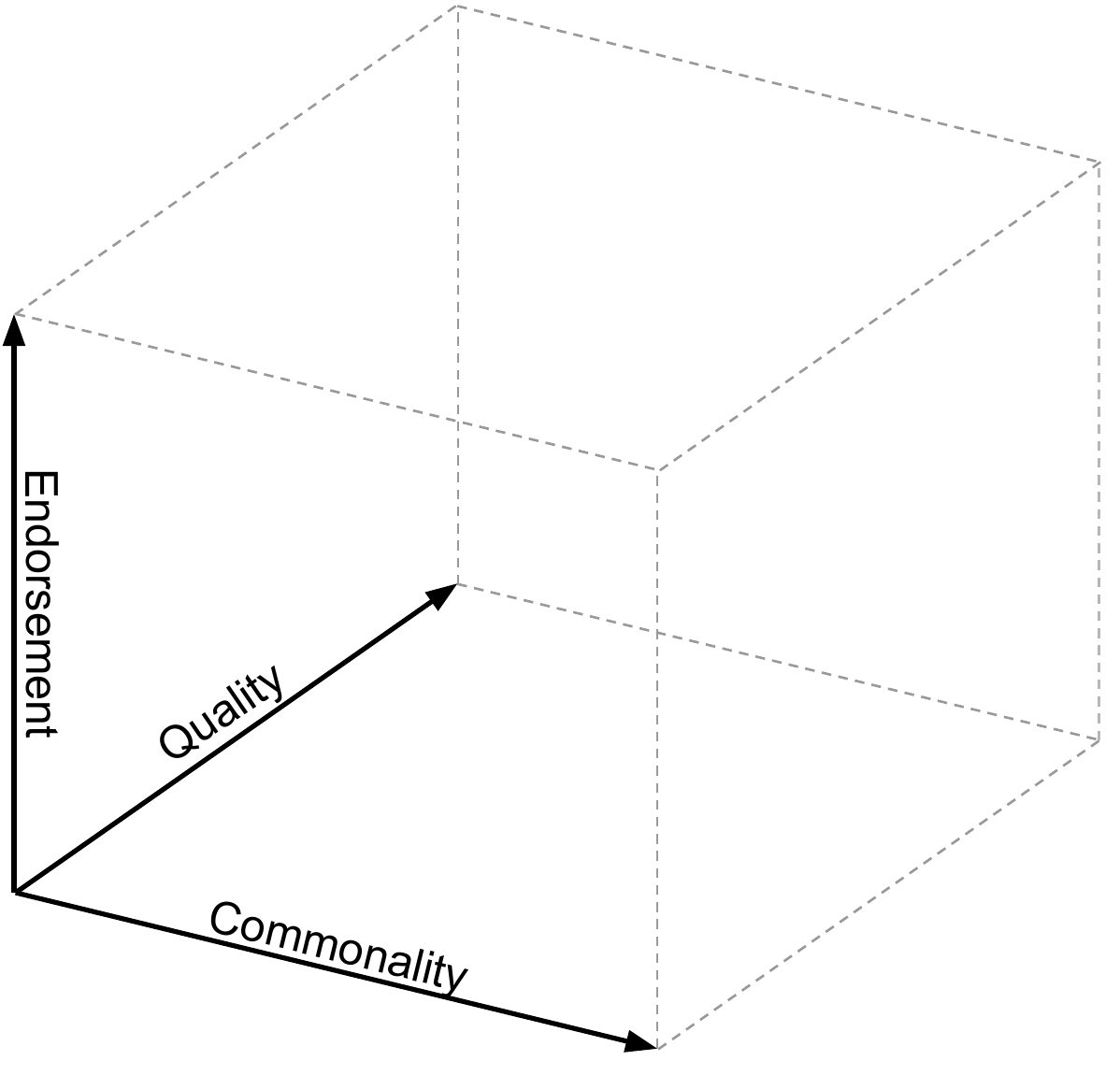}
        \caption{A model that categorically distinguishes practice-related characteristics.}
        \label{fig:3axes}
    \end{figure}

    From this, we envision a 3-D graph with axes of quality (good, best, \"{u}ber), frequency of use or commonality (common practice below requirement), and the level of endorsement a practice has received (formal standardization, recommendation, guideline). See Figure~\ref{fig:3axes}. The three-axis model provides a mental image of how one piece of advice compares to another within three-dimensional space. This visualization helps us conceptualize more easily than simply comparing two pieces of advice based on their wording, and may suggest relationships between items placed within this space.
    
%%%%%%%%%%%%%%%%%%%%%%%%%%%%%%%%%%%%%%%%%%%%%%%%%%%%%%%%%%%%%%%%%%%%%%%%%%%%%%%%%%%%%%%%%%%%%%%%%%%%
\section{Lifecycle of IoT Devices}
\label{sec:lifecycle}

In this section, we present a simplified model of the lifecycle of an IoT device. We discuss the major and minor stages a device moves through during its life. This allows an understanding of the relationship between an IoT device manufacturer and the end-users that will be impacted by the manufacturer's security-related design choices. This will aid later analysis of the roles these and other stakeholders play in security.

\begin{figure}[t!]
    \centering
    \includegraphics[width=0.48\textwidth]{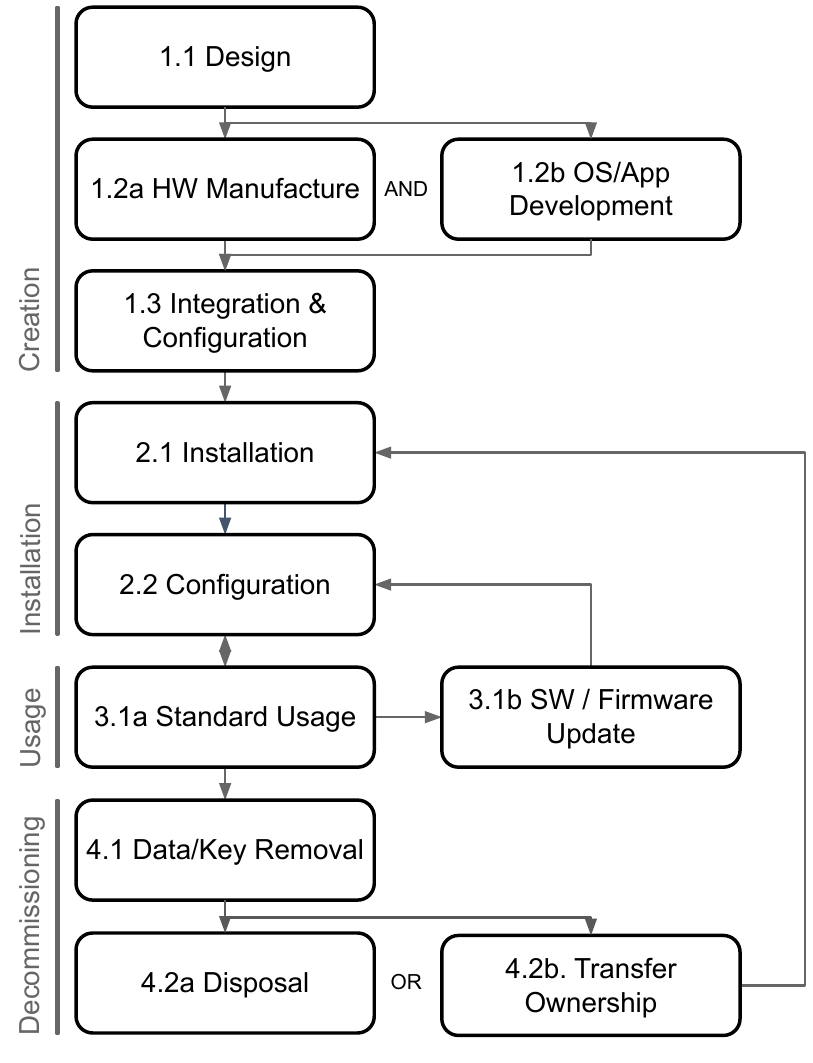}
    \caption{A typical lifecycle for an IoT device, from initial design to end-of-life.}
    \label{fig:lifecycle}
\end{figure}

The lifecycle of an IoT device includes all major phases a device could go through from its early design up to the time it is discarded (disposed-of, or otherwise never used again)\cite{Garcia-Morchon2019_2}. This plays a significant role in the discussion surrounding best practices, particularly for IoT devices, as their nature is to be long-lasting, fairly idle, and more commonly link our digital environments to our physical environments. Decisions made within one part of the lifecycle may affect later phases. Once these products have left the hands of manufacturers, it becomes more challenging to properly address vulnerabilities due to the more highly resource-constrained nature of IoT products. It is important that we understand what processes take place within each major phase. Figure~\ref{fig:lifecycle} presents the typical lifecycle of an IoT device and the major phases a device is expected to go through. These phases are discussed here.

    \subsection*{1) Creation}%%%%%%%%%%%%%%%%%%%%%%%%%%%
    \label{phase:creation}
    The Creation phase consists of four sub-phases that all take place at the purview of the manufacturer. This is where a product is designed, developed, and otherwise readied for consumption by a user. The key feature that separates the Creation phase from all other phases is that it takes place before the end-user is involved. Sub-phases may be done with the end-users in mind, but the product has not yet been received by them.
    
        \paragraph*{1.1) Design}
        The Design phase is where the design of both the hardware and software of an IoT device is done. This includes design considerations for security such as choice of software toolkits or libraries, selecting operating systems, or design of software that will run on a device. This phase concludes when the designs have been formalized and ready for manufacturing of the product. This includes the design of any hardware and software (OS, IoT app) if these components are not sourced from entities outside of the manufacturer. It should be noted that designs may change over time and potentially require updates for future models, so the design phase may be mentally considered to be more cyclical; however, in our vision of the lifecycle, we consider a singular model/version of a device.
        
        \paragraph*{1.2a) Hardware Manufacture}
        The Hardware Manufacture phase is where the physical device is produced by the IoT device manufacturer (if done in-house) or dedicated hardware manufacturer. While this phase primarily concerns the development of hardware, as firmware (``software that is embedded in a hardware device'' \cite{Costin2014}) is so closely tied to hardware, firmware is developed in this phase. This phase concludes when a device is physically complete and ready to have its software loaded and preconfigured.
        
        \paragraph*{1.2b) OS/App Development}
        The OS/App Development phase takes place in a conceptual parallel to the Hardware Manufacture phase. This is where any operating system or core-functionality software is built. The OS may be developed by the product manufacturer, or at another time by an external organization (e.g., a Linux distribution is likely not developed by an IoT manufacturer, but may be used by them). The core-functionality software is the software that is required to provide a device with the ability to undertake its standard function, e.g., the software that will control the camera in an IP camera, or control temperature in a thermostat. This phase concludes when the OS and software are ready to be integrated with the hardware.
        
        \paragraph*{1.3) Product Integration \& Preconfiguration}
        The Product Integration \& Preconfiguration phase is where software is loaded onto an IoT device's hardware and configured with any settings or data that will be required for basic functionality and/or security process. This included the application of an OS or any software to hardware, core functionality applications onto an OS, provisioning of cryptographic material (e.g., key pairs, long-term keys, certificates, trust-anchors), or anything else that must be configured before being used by the end-user. In the case of a device with very basic functionality such as a microcontroller with minimal functionality and no non-firmware software to run, this step may not be required. This phase concludes when the device is in a ready-to-ship state (i.e., ready to be sent to a retailer or customer). There may exist an additional packaging phase after this where the product is put into its retail packaging, but having no technical implications, we do not consider this within our model.
    
    \subsection*{2) Installation}%%%%%%%%%%%%%%%%%%%%%%%%%%%
    \label{phase:installation}
    The Installation phase consists of two sub-phases where the user has received the product, but has not yet readied it for standard usage. These two sub-phases consist of what is commonly referred to as ``onboarding'' or ``bootstrapping'' (often used interchangeably or meaning slightly different things, depending on who uses it)\cite{IDBootstrap} where a number important technical details such as key management, identification, and trust relationships are established; and more general user-focused configuration. 
        
        \paragraph*{2.1) Installation}
        The Installation sub-phase is when a user installs the device in its environment.  Examples of this may be the placement of the device with connection to a power source or have a battery installed/activated (if not already active), connecting to physical network, and powering the device on. While physical connection to a network (i.e., inserting network cable) is considered in the installation phase, any networking configuration to do with the physical network connection or wireless connection takes place in the Configuration phase. This phase's sole purpose is to place a device into an environment and prepare it for user configuration and concludes when this has been done.
        
        \paragraph*{2.2) Configuration}
        The Configuration phase is where a user provides any configuration to the device that they need in order to set the device into its standard usage state, and provide any security configuration for standard usage. This stage includes processes such as network configuration (particularly in the case of wireless communication, as the wired connection is done in the previous phase and commonly automatically configured by various networking protocols), the setting or resetting of passwords for device access, the generation of new keying material and onboarding (if not pre-generated by the manufacturer or requiring new key relationships), or setting any functional/behavioral settings that the user may want from their device. More drastically, this phase may also include performing a factory reset of the device, requiring that users fully reconfigure their device. This phase concludes when the device has been configured to the end-user's liking and any onboarding/bootstrapping functions have been completed.

    \subsection*{3) Usage}%%%%%%%%%%%%%%%%%%%%%%%%%%%
    \label{phase:usage}
        
        \paragraph*{3.1a) Standard Usage}
        The Standard Usage phase is where the device is functioning as intended (e.g., a light bulb provides light, a smart thermostat controls temperature, a home security camera provides live camera access to users). This phase concludes when a software/firmware update is available for the device (and the user chooses to update), when the user wishes to make configuration changes to the device (moving back to the Configuration phase), or when the user wishes to discontinue the device's usage. While seemingly small within our model, this is the phase where the device is expected to spend the majority of its life.
        
        \paragraph*{3.1b) Software/Firmware Update}
        The Software/Firmware Update phase is where the device receives and installs new software. This phase concludes when a device has successfully updated, or fails to update. In the case of a successful update, reconfiguration may be required for any settings that may have been reset, or for new functionality to be configured. In the case of an update failure, the device may fall back to its standard usage or cease to function and require disposal.

    \subsection*{4) Decommissioning}%%%%%%%%%%%%%%%%%%%%%%%%%%%
    \label{phase:decommissioning}
    
        \paragraph*{4.1) Data/Key Removal}
        The Data/Key Removal phase is where sensitive data is removed from a device prior to being removed from an environment. If private user data is not removed, future users or attackers may be able to recover data. Keying material is removed to prevent session keys, network connection information, or other sensitive data from being leaked upon future inspection. This may also include the removal of data stored by manufacturers or service providers in other devices or in cloud servers. This phase may also be referred to as ``offboarding'', where many of the tasks completed during onboarding are undone. This phase concludes once all specified data has been removed, or if a user chooses to skip the removal.
        
        \paragraph*{4.2a) Disposal}
        The Disposal phase is the final phase of a device's lifecycle if ownership is not transferred. The device is uninstalled and discarded in some way, whether it is through simple garbage disposal or recycling of the device's parts.
        
        \paragraph*{4.2b) Transfer Ownership}
        The Transfer Ownership is the terminating phase in a single owner's IoT device lifecycle, where a device's ownership is transferred (whether sold or gifted) to another end-user or organization. The device is uninstalled from the environment. Upon receiving the new device, the new owner will have to return to the Installation phase in the device's lifecycle to begin as the original owner did, or if configuration still exists, the new owner may be able to immediately begin usage.
        
%%%%%%%%%%%%%%%%%%%%%%%%%%%%%%%%%%%%%%%%%%%%%%%%%%%%%%%%%%%%%%%%%%%%%%%%%%%%%%%%%%%%%%%%%%%%%%%%%%%%
\section{Analysis and Categorization of Practices}
\label{sec:categorization}

In this section we discuss a large collection of industry, academic, and government security advice. We categorize the items in this collection based on when they are implemented on the IoT lifecycle (Figure~\ref{fig:lifecycle}), and whether or not each item aligns with our definition of a ``best practice'', versus an outcome (type-V or type-S). Noting where practices are implemented throughout the IoT lifecycle allows us to see which stakeholders have the greatest impact on overall device security. Matching our definition of ``best practice'' with each item allows us to get a general sense of how well existing advice lists and literature have provided actionable practices.

The UK Department for Digital, Culture, Media \& Sport (DCMS) has published a collection of IoT security recommendations, standards, and guidelines \cite{DCMS1}. These were extracted from academic, industry, and government documents for manufacturers of IoT products \cite{DCMS1}, and includes 1014 practices (after explicit duplicates were removed) collected from 69 documents, from 49 different organizations. They categorize each practice into one of 13 ``outcome focused guidelines'' \cite{DCMS2} (also in the process of being adopted by the government of Australia \cite{AUSIoT}). A few examples of their guidelines includes ``no default passwords'', ``communicate securely'', and ``make it easy for customers to delete personal data''. Table~\ref{tab:practices} provides a full listing of the 13 guidelines.

We note that the UK document is entitled ``\textit{Code of Practice} for Consumer IoT Security'' \cite{DCMS2}. We consider a code of practice as a regulation (e.g., residential/commercial building code, electrical code, fire code, etc.), resulting in our classification as a requirement in Table~\ref{tab:qualifyingterms}; however, their document outlines ``outcome focused \textit{guidelines}''. We classify guidelines under the formal recognition category. As discussed in Section~\ref{sec:bestpractices}, we place requirements and formal recognition on their own continuum (endorsement).

While this is a significant collection of industry work and it does specifically target stakeholders that would implement practices, their categorization does not focus on the lifecycle of IoT products. This is an important consideration when discussing IoT as, given the long-lasting and ubiquitous nature of the devices, significant numbers of practices must be implemented for different stages of a product's life (Figure~\ref{fig:lifecycle}). Different stakeholders will be affected depending on the practice and when it is implemented, often putting distance between the potential lack of practice implementation and the impact of such a decision. Further, upon inspection of DCMS collected best practices, we found the majority of practices can belong to multiple lifecycle stages. Considerations for lifecycle stage and stakeholder impact expands on the DCMS' categorization \cite{DCMS2}.

\begin{figure}[t!]
    \centering
    \includegraphics[width=0.48\textwidth]{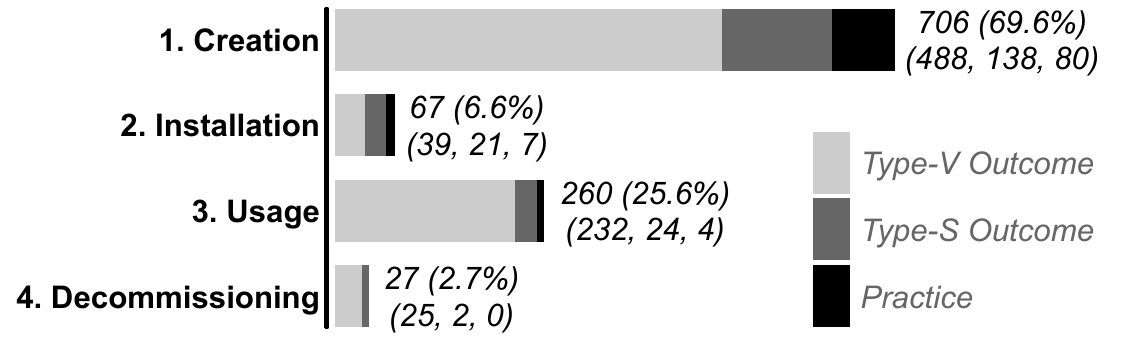}
    \caption{Distribution of the listed 1014 items (guidelines, recommendations, best practices) across IoT major lifecycle phases. Total percentage exceeds 100 as some practices appear in multiple phases. (X, Y, Z) denotes the respective number of practices for (type-V outcomes, type-S outcomes, practices). Bars of small numbers have had their widths slightly exaggerated to improve readability.}
    \label{fig:representation}
\end{figure}

%METHODOLOGY
Regarding methodology, we manually categorized each advice item (1014 items) in two ways: where in the lifecycle each item would be best implemented from a security perspective, and if an item was an outcome or a practice. For the first lifecycle tagging, each item was associated with which phase in the IoT lifecycle they belong to. This was done by examining each item and determining which stakeholder would be the most appropriate to implement it, and matching where their work was done to the lifecycle. For the second categorization, each item was manually tagged as being a type-V outcome, type-S outcome, or practice. If an item met our working definition of a practice, it was tagged as a ``practice''. If it was not a practice, the level of specificity/vagueness was judged: items that were highly vague were tagged as a type-V outcome, items that were more specific and implying actions were tagged as type-S outcomes. There were a number of items in the collection that were not goals to be reached, but often just descriptions of techniques and technologies. In these cases, it was assumed that they are meant to be targets to achieve, and we tagged these as type-V outcomes.

Figure~\ref{fig:representation} highlights our results categorizing each major phase in the IoT lifecycle (Figure~\ref{fig:lifecycle}) and the number of practices from the DCMS collection that would be implemented in each. The practice distribution among phases paints an important picture for the general implementation of best practices: the overwhelming majority (about 69.6\%) of practices relate to stages before leaving the manufacturer's hands (the Creation phase), indicating that it is the designers and manufacturers of a product that are the parties in a position to implement them. This is a rather apparent conclusion to draw from these results, but it quantifies the significance and importance of the security posture developed during this stage.

Figure~\ref{fig:representation} also highlights our second categorization of whether or not an item is an outcome or a practice. We find that 784 of the 1014 recommendations (77.3\%) are vague and not practices at all (type-V outcomes), 185 of the 1014 (18.2\%) specify outcomes that are somewhat more specific and imply actions, do not suggest actual practices (type-S outcomes); and only 91 of the 1014 (9\%) are actual practices per our working definition. This second categorization shows that organizations---often highly credible ones---are producing recommendations for manufacturers that are not, in our view, actionable, thus appear unlikely to be properly implemented. This may require a significant shift in how we produce best practices if manufacturers are expected to follow recommendations.

We divide the stakeholders of IoT security into two groups: ``pre-sale'' stakeholders that are involved during the Creation phase prior to being sold, and ``post-sale'' stakeholders that are affected by the decisions of ``pre-sale'' stakeholders. Note that this ``sale'' describes when an end-user purchases the product from a retailer, not when a user purchases the product from another end-user or some other pre-owned method. The decisions made in this phase have a direct impact on post-sale stakeholders (e.g., users, platform providers, app developers, OS developers) as weakness in product security posture may have significant consequences later in the lifecycle.

%%%%%%%%%%%%%%%%%%%%%%%%%%%%%%%%%%%%%%%%%%%%%%%%%%%%%%%%%%%%%%%%%%%%%%%%%%%%%%%%%%%%%%%%%%%%%%%%%%%%
\section{Primary IoT Security Stakeholders}
\label{sec:stakeholders}

In this section we consider the major stakeholders involved in the consumer IoT lifecycle. We discuss the roles and responsibilities of stakeholders, and the impact they have on the IoT lifecycle and security posture of devices produced by manufacturers. Understanding each stakeholder's impact for the various components and features of a device allow us to more clearly see how there are many moving parts that impact a security posture.

\begin{table}
    \caption{Stakeholders impacting the lifecycle of an IoT Device.\hspace{\textwidth}Herein, our focus is primary stakeholders.}
    \centering
    \label{tab:stakeholders}
    \begin{tabular}{@{}lccrc@{}}
        \toprule
        & Pre-sale   & Post-sale         \\  \midrule
        %%%%%%%%%%%%%%%%%%%%%%%%%%%%%%%%%%%%%%%%%%%%%%%%%%%%%%%%%%%%
        Primary Stakeholders              &   &   \\  
        \cmidrule{1-1}   
        HM: IoT hardware provider         & \checkmark &   \\    
        OD: IoT OS provider               & \checkmark & \checkmark \\    
        AD: IoT software/app developer    & \checkmark & \checkmark \\     
        PI: IoT product integrator    & \checkmark &   \\     
        DM: IoT device manufacturer       & \checkmark &   \\     
        EU: End-user                      &   & \checkmark \\   \midrule
        %%%%%%%%%%%%%%%%%%%%%%%%%%%%%%%%%%%%%%%%%%%%%%%%%%%%%%%%%%%%
        Secondary/Other Stakeholders&&            \\   
        \cmidrule{1-1}   
        Smartphone (OS, app) developers   &   & \checkmark \\     
        Smartphone app store host         &   & \checkmark \\    
        Third-party cloud platform        &   & \checkmark \\    
        Third-party cloud host            &   & \checkmark \\     
        Third-party cloud app developer   &   & \checkmark \\    
        Government regulators             & \checkmark &   \\    
        Standards and protocol designers  & \checkmark &   \\  
        \bottomrule 
    \end{tabular}
\end{table}

For any given product there are numerous stakeholders. Depending on the product, some stakeholders are more impacted by security issues than others. Each stakeholder has a number of goals that dictate their actions with regards to their role within the IoT lifecycle and how they are affected by the implementation of best practices, and consequences that will affect them if the practices are not implemented. It is important to note the major stakeholders as each pre-sale stakeholder may be responsible for implementing different practices based on their role in the production of a device. These stakeholders are discussed here, and Table~\ref{tab:stakeholders} lists these stakeholders for reference.

    \subsection{Pre-Sale Stakeholders}
        
        \paragraph*{IoT Hardware Provider}%%%%%%%%%%%%%%%%%%%%%%%%%%%%%%%%%%%%%%%%%%%%%
        The hardware provider specifically manufacturers the hardware used in an IoT device. On the high, less constrained end of hardware, an example is the Raspberry Pi 4 \cite{rPi4} or Arduino Due \cite{arduinoDue}. On the low, more highly-constrained end of hardware, it is becoming more difficult to find pre-built (containing a processor of some sort, memory, I/O) boards that fall within RFC 7228's definition of a highly-constrained (Class 0) device \cite{rfc7228}.
        
        \paragraph*{IoT OS Provider}%%%%%%%%%%%%%%%%%%%%%%%%%%%%%%%%%%%%%%%%%%%%%
        The IoT OS provider stakeholder is responsible for developing the OS for an IoT device. For the most part, this only applies to devices that are capable of running some form of OS rather than the highly constrained devices that run only the most basic firmware and application-specific software; however, a wide variety of hardware exists \cite{Hahm2016}. Examples in this space include such OSs as RIOT \cite{Baccelli2018}, Contiki \cite{ContikiNG}\cite{Dunkels2004}, or FreeRTOS \cite{freeRTOS}. The IoT OS provider writes only the OS and surrounding functionality. The OS would be applied to hardware, and any IoT apps would be installed into the OS for application-specific functionality.
        
        \paragraph*{IoT Software/App Developer}%%%%%%%%%%%%%%%%%%%%%%%%%%%%%%%%%%%%%%%%%%%%%
        IoT software/app developers are those that develop the core software that an IoT device will run. This development is intended to run as software on top of an IoT OS, or as the near-metal firmware running on devices without OSs. As a fairly broad example, an ``IoT software/app'' might be the software that collects, analyses, and reports temperature readings on a smart thermometer device, not the underlying software that provides the basic OS functions and communication. Quite often these are developed by the manufacturer (or contracted by the manufacturer), as they are required to be very application-specific.
        
        \paragraph*{IoT Product Integrator}%%%%%%%%%%%%%%%%%%%%%%%%%%%%%%%%%%%%%%%%%%%%%
        The product integrator is the entity that receives hardware from the hardware manufacturer and software from the software developer (IoT OS, apps), and integrates them into the final product. This may include installing IoT apps onto an OS, then loading the resulting image onto the hardware. The product integrator may be responsible for preconfiguring a device (i.e., setting base configuration or loading crypto materials such as keys, certificates, etc.), or they may send the devices back to the manufacturer for this process. Product integrators combine any hardware and software to complete the basic IoT product that a manufacturer will later sell. They are not necessarily interested in the implementation of security features and best practices; however, their role may be to apply certain pieces of information or installing features that assist with this.
        
        \paragraph*{IoT Device Manufacturer}%%%%%%%%%%%%%%%%%%%%%%%%%%%%%%%%%%%%%%%%%%%%%
        The IoT device manufacturers design the product, manufacturer it, and ship it to users or retailers for distribution to users. They may be responsible for the development of software or hardware, but this is commonly done by another, contractually-obligated stakeholder. In some cases, the device manufacturer may be responsible for integrating any software components with hardware (acting as the product integrator), or specific portions of the full process (design, hardware manufacturing, software development, warehousing, shipping).
        
    \subsection{Post-sale Stakeholders}
    Post-sale stakeholders are those that are primarily concerned with an IoT after being purchased by a user (and including the users). These stakeholders have little impact on the decisions made pre-sale during the Creation phase. This makes them far more impacted by security vulnerabilities caused by the lack of best-practice implementation before the sale was made. For this work, we focus primarily on the end-user, as they are the ones purchasing devices, using services, and being the most directly impacted by upstream problems.
    
        \paragraph*{End-users}%%%%%%%%%%%%%%%%%%%%%%%%%%%%%%%%%%%%%%%%%%%%%
        The end-users are the individuals or organizations that purchase IoT products to install in their homes or businesses. End-users---especially for more consumer-grade products and/or services---are directly impacted by any functionality changes; or indirectly via back-end decisions that may affect their usage of the product/service. End-users essentially have only one goal: use the product they purchased for its intended function, for the time that they want to use it for. This means they must be able to successfully move through all post-Creation lifecycle phases. Disruption events such as service down-time, malfunction, malicious attacks that hinder their ability to use the device, or some other event that the user deems out of their control will be a problem for them. For the users, the greatest consequence of manufacturers not implementing best practices is the potential vulnerabilities that may exist in their device. The greatest influence the end-users have on the manufacturers is with their wallets. In theory, if enough end-users refused to purchase a manufacturer's products or services, sales would be lower and manufacturers may find themselves adjusting to fit the needs of the end-users. Besides their purchasing power, end-users have little influence on manufacturers. End-users also have influence over a number of other primary and secondary stakeholders. Their usage of a product directly affects the usage of related products and/or features such as the platforms and services provided (third-party platform providers), operating systems (OS developers), or software running on or around a device (app developers).
        
    \subsection{Secondary/Other Stakeholders}
    Beyond the primary pre- and post-sale stakeholders, there are a number of other groups that are impacted. While not being the subject of focus here as a ``primary'' stakeholder, these are the stakeholders that are likely to be affected by the major stakeholders and decisions made pre-sale. As it highlights the breadth of potential impact for security best practices, we make note of some secondary/other stakeholders in Table~\ref{tab:stakeholders}, but this is not an exhaustive list. The number of secondary stakeholders highlights further impact that vulnerabilities and problems may have on other post-sale stakeholders other than end-users.

%%%%%%%%%%%%%%%%%%%%%%%%%%%%%%%%%%%%%%%%%%%%%%%%%%%%%%%%%%%%%%%%%%%%%%%%%%%%%%%%%%%%%%%%%%%%%%%%%%%%
\section{Motivating Better Security Posture}
In this section we discuss possible methods to motivate pre-sale stakeholders to adopt best practices to improve the security posture for their devices. Understanding how challenging it may be to motivate pre-sale stakeholders to adopt this better security posture highlights the importance of any means to reduce adoption resistance including providing clear, actionable best practices.

While some methods exist for post-sale stakeholders to provide additional security for their devices, ensuring stronger security practice implementation requires significant cooperation with the pre-sale stakeholders. Expecting the IoT industry to self-regulate has not worked so far, and with the scale that IoT is expected to approach, strategies need to be put into place as soon as possible rather than relying on the self regulation to take place when the stakes have increased.

Manufacturers are not well-incentivized to implement the established best practices. No incentive for implementation may result in no implementation, as a cost-benefit analysis may reveal an implementation is not worth the cost. This issue may be exacerbated by the scale of a company looking to produce IoT devices; smaller companies may not be able to introduce practices that require more significant resource investment. Further exacerbating this issue is the aspect of how reasonable a practice is to be implemented, which ties directly into the cost-benefit analysis. Different practices may be more reasonable for some stakeholders to implement than others (e.g., large manufacturers may have the means---although not necessarily the motivation---to implement more costly practices).

%End-user pressure
Perhaps the most natural approach is to rely upon end-users to police poorly secured devices via the ``vote with your wallet'' motto. In a perfect world, poorly secured devices would not be purchased, but this does not work in reality for a number of reasons:
\begin{itemize}
    \item Awareness of issues---The end-users must be informed of the security issues that plague a product, but this is often difficult to find. The average user would not be searching academic or industry research citing weaknesses of certain products or features, and are less likely to stay up-to-date on security- or tech-related websites and blogs.
    \item End-user resolve---The end-users must maintain their resolve to not buy products if their security in inadequate. This requires high levels of end-user education about existing problems with devices (see previous point). When so many products have high usability, users may opt for the functionality of the device (very visible and tangible to the user) over the security, which tends to be more invisible to them. Users will be drawn to the products that work as they want them to work, following the ``it just works'' mantra. Alternatively, if a product is widely purchased, consumers may feel pressured to purchase it based on social pressure, and ignore security issues.
\end{itemize}

Another approach is to continue with the current mechanism for change via academic/industrial publications about vulnerabilities. This provides visibility for the larger vulnerabilities (e.g., ``IoT Goes Nuclear'' \cite{Ronen2017}, Mirai botnet \cite{Antonakakis2017}), but the smaller issues that should still be addressed may not provide enough pressure on the manufacturer.

When we say we want to provide ``incentive'', we really mean that we are looking to add pressure to manufacturers. Being incentivized implies a positive reason to change, while being pressured has a negative connotation. As a business will act in its best interest (the cost-benefit analysis problems from above), more concrete and perhaps drastic methods are required. One method already in limited use is government regulation. While fraught with concern for overreach, lengthy enactment periods, and inadequacy, government regulation has the ability to enforce certain practices within a region. Examples of this being done include the Government of California (USA), which has passed a bill requiring IoT device manufacturers to provide ``reasonable security features'' \cite{Bill327}; and the European Union's General Data Protection Regulation (GDPR), which provides guidelines and regulation for European service providers regarding user data usage \cite{GDPR2016}.
    
While using governments to strictly enact and enforce regulation on companies is likely to have a more concrete impact on manufacturers, it has a number of drawbacks that need to be taken into account:
\begin{itemize}
    \item Government overreach: Asking a government to interfere in an industry's actions may seem rather heavy-handed, and to many, an overreach of government power where the government should not be interfering in the workings of an industry. This view changes from country to country where some view the role of government as more important than others. For certain industries such as healthcare, it may be more understandable to some to be highly regulated by governments, as healthcare outcomes are directly responsible for the physical well-being of its citizens. Tech industries are not (yet) so directly tied to the well-being of the citizens of a country, and issues within the industry have historically not been critical to the safety of a country's populace. This is, however, changing as IoT products become more intertwined with our daily life, with some even being responsible for our health. As this happens, government regulation may seem like more of a logical necessity, as healthcare is seen in many countries.
    \item Government vs. Industry/Academic Disagreement: There may exist the problem where those writing regulation disagree with industry or academic researchers (or vice-versa) about what practices should be enforced. This may lead to inadequate practices being forced on manufacturers and adopted due to political pressure versus sound technical reasoning and consensus.
\end{itemize}

This said, government regulation has a number of benefits that should be considered:
\begin{itemize}
    \item Localized implementation---Attempting to have regulation enacted across borders is challenging; however, regulation within a single government may be much more approachable as a starting point and allowing more global momentum to build.
    \item Change for one, change for all---A manufacturer that has to produce two versions of the same product (one for one government market, another for the rest of the world) may opt to simply roll all improved security versions into the same product and supply it to the rest of their markets. This way, stakeholders from markets unaffected by government regulation may reap the benefits of the additional security without having to go through the process of regulation.
\end{itemize}

Legislative bodies of significant world markets have a greater ability to make government regulation be effective and enforce change as the manufacturer will likely not want to abandon such a large market to sell in. Smaller bodies do not have this opportunity. Alternatively, governments and large organization can enact change via their purchasing power. These entities that purchase products for wide internal use can put restrictions on their orders where the product will only be purchased if it conforms with specific requirements. For example, if a smart light bulb was planned to be used across all government buildings within a country, stating that they will not purchase the product if it is vulnerable to attack X, Y, or Z may be enough to motivate manufacturers to fix the related vulnerabilities. Much like the above, this only works for entities with great buying power.

Despite the above drawbacks, government regulation appears to be the most effective---while quite heavy-handed---way to most strictly enforce change within a region, as it threatens manufacturers' abilities to sell within a market, which the other avenues have a more difficult time achieving. With this said, not all practices can be (or necessarily should be) candidates for government regulation. This concept is discussed further in the next section.  

%%%%%%%%%%%%%%%%%%%%%%%%%%%%%%%%%%%%%%%%%%%%%%%%%%%%%%%%%%%%%%%%%%%%%%%%%%%%%%%%%%%%%%%%%%%%%%%%%%%%
\section{Implementation of Best Practices}
\begin{figure}[t!]
    \centering
    \includegraphics[width=0.48\textwidth]{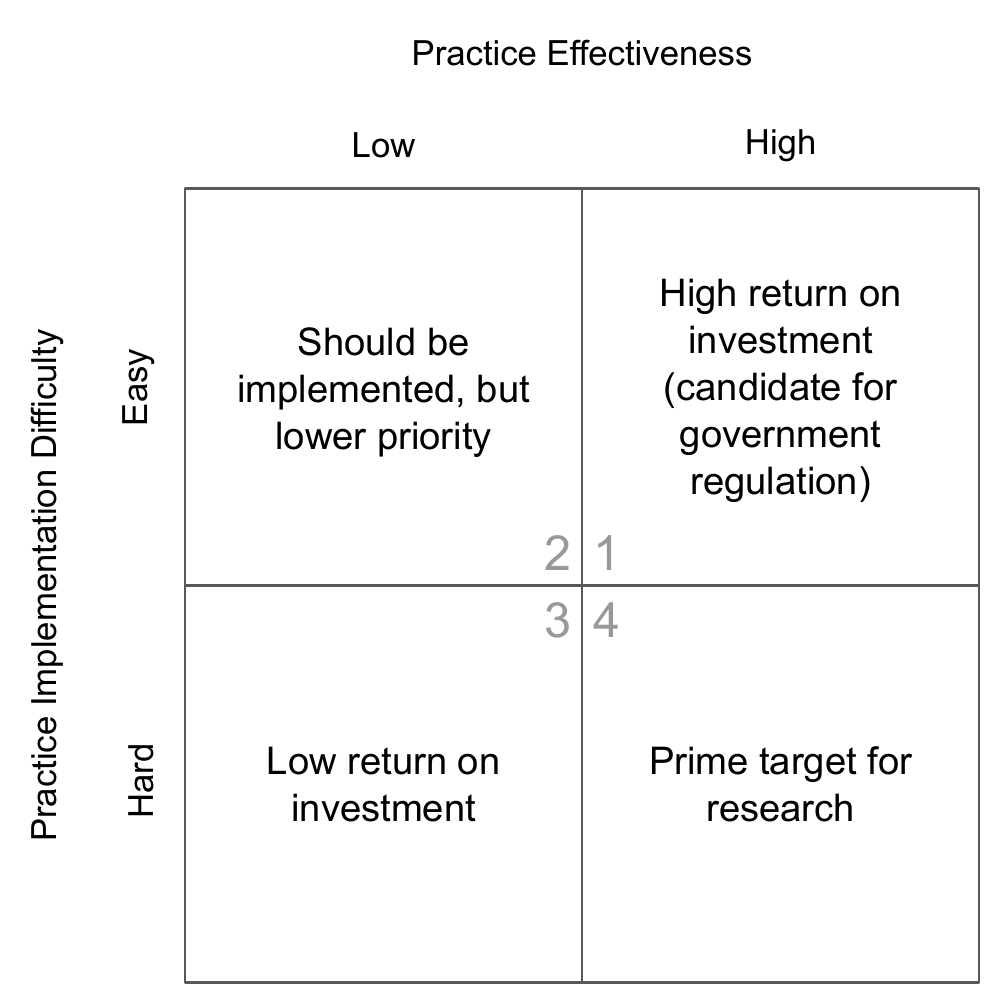}
    \caption{Four-quadrant implementation difficulty vs. practice effectiveness grid. Table~\ref{tab:guidelineupdates} gives examples for each quadrant.}
    \label{fig:grid}
\end{figure}

In this section, we map the UK's 13 guidelines (for improving the security of IoT products, introduced in Section~\ref{sec:categorization}) to where they exist within the IoT device lifecycle of Figure~\ref{fig:lifecycle}, and outline which pre-sale stakeholders are in the best position to implement them. These ``outcome focused guidelines'' \cite{DCMS2} are comprised of a wide variety of standards, guidelines, and recommendations from various academic, industrial, and government institutions, which were then aggregated to form the guidelines. This mapping highlights again how a variety of major pre-sale stakeholders are required to achieve these outcomes. For each of the 13 guidelines, we reinterpret them into new type-S outcomes that are more specific. We propose a system (using Figure~\ref{fig:grid}) for categorizing practices by their difficulty of implementation and effectiveness in improving security for a system, and map these reinterpreted outcomes based on this system. This is done to identify which existing outcomes are insufficient for improving IoT security posture based on their implementation difficulty, or their lack of effectiveness in improving security.

For pre-sale stakeholders looking to potentially implement a best practice, the ideal practice would be both easy to implement (i.e., does not require extreme amounts of time, resources, or manpower) and highly effective (i.e., provides a significant boost to security posture). Figure~\ref{fig:grid} gives a simple 2x2 grid of combinations for the difficulty of implementing any given practice, and its security effectiveness. Any candidate currently falling into quadrant 4 is a prime candidate for security research, in order to find a new (less difficult) technical mechanism, ideally even moving into quadrant 2.

\begin{table*}[t]
    \centering
    \caption{Analysis of UK DCMS guidelines for consumer IoT security. Table indicates stakeholders in a position to implement each, and where in the lifecycle they are implemented. ``\checkmark'' denotes a type-S outcome, which suggests an action (Section \ref{sec:background/actionable} defines ``action'' and ``outcome''). \textit{DM} (Device Manufacturer). \textit{HM} (HW Manufacturer). \textit{OD} (OS Developer). \textit{AD} (App Developer). \textit{PI} (Product Integrator). End-users do not appear in the stakeholder column as these guidelines are specifically designed for pre-sale stakeholders.}
    \label{tab:practices}
    \begin{tabular}{@{}p{7.0cm}p{2cm}p{1.00cm}p{2.0cm}p{1.75cm}@{}}
        \toprule
        UK Guideline \cite{DCMS2} & Example References & Implies Action? & Responsible Stakeholder & Lifecycle Stage (from Fig. \ref{fig:lifecycle}) \\ \midrule
        %%%%%%%%%%%%%%%%%%%%%%%%%%%%%%%%%%%%%%%%%%%%%%%%%%%%%%%%%%%%%%%%%%%%%%%%%%%%%
        UK-1. ``No default passwords'' & 
        \cite{ENISA1}, \cite{CSA2}, \cite{IEEE1} & 
        \hfil\checkmark & %Actionable?
        PI & 
        1.3 \\ \addlinespace[0.05cm]
        %%%%%%%%%%%%%%%%%%%%%%%%%%%%%%%%%%%%%%%%%%%%%%%%%%%%%%%%%%%%%%%%%%%%%%%%%%%%%
        UK-2. ``Implement a vulnerability disclosure policy'' & 
        \cite{ACSCSS1}, \cite{BITAG1}, \cite{ENISA1} & 
        \hfil\checkmark & %Actionable?
        DM & 
        1.1, 3.1a \\ \addlinespace[0.05cm]
        %%%%%%%%%%%%%%%%%%%%%%%%%%%%%%%%%%%%%%%%%%%%%%%%%%%%%%%%%%%%%%%%%%%%%%%%%%%%%
        UK-3. ``Keep software updated'' & 
        \cite{BITAG1}, \cite{CableLabs1}, \cite{AIOTI1} & 
        & %Actionable?
        DM, OD, AD & 
        1.1, 1.2b, 3.1b \\ \addlinespace[0.05cm]
        %%%%%%%%%%%%%%%%%%%%%%%%%%%%%%%%%%%%%%%%%%%%%%%%%%%%%%%%%%%%%%%%%%%%%%%%%%%%%
        UK-4. ``Securely store credentials and security-sensitive data'' & 
        \cite{ENISA1}, \cite{CSA1}, \cite{CSA2}  & 
        \hfil & %Actionable?
        OD, HM, PI & 
        1.2a, 1.2b, 1.3 \\ \addlinespace[0.05cm]
        %%%%%%%%%%%%%%%%%%%%%%%%%%%%%%%%%%%%%%%%%%%%%%%%%%%%%%%%%%%%%%%%%%%%%%%%%%%%%
        UK-5. ``Communicate securely'' & 
        \cite{BITAG1}, \cite{ENISA1}, \cite{CSA1}  & 
        \hfil & %Actionable?
        OD, AD & 
        1.2b, 1.3 \\ \addlinespace[0.05cm]
        %%%%%%%%%%%%%%%%%%%%%%%%%%%%%%%%%%%%%%%%%%%%%%%%%%%%%%%%%%%%%%%%%%%%%%%%%%%%%
        UK-6. ``Minimise exposed attack surfaces'' & 
        \cite{AIOTI3}, \cite{ATT1}, \cite{NYC2} & 
        & %Actionable?
        HM, OD, AD & 
        1.2a, 1.2b, 1.3 \\ \addlinespace[0.05cm]
        %%%%%%%%%%%%%%%%%%%%%%%%%%%%%%%%%%%%%%%%%%%%%%%%%%%%%%%%%%%%%%%%%%%%%%%%%%%%%
        UK-7. ``Ensure software integrity'' & 
        \cite{ENISA1}, \cite{IIC1}, \cite{GSMA4} & 
        & %Actionable?
        OD, AD & 
        1.2b \\ \addlinespace[0.05cm]
        %%%%%%%%%%%%%%%%%%%%%%%%%%%%%%%%%%%%%%%%%%%%%%%%%%%%%%%%%%%%%%%%%%%%%%%%%%%%%
        UK-8. ``Ensure that personal data is protected'' &
        \cite{AIOTI3}, \cite{AIOTI4}, \cite{NYC1} & 
        \hfil & %Actionable?
        OD, AD & 
        1.2a, 1.2b, 1.3 \\ \addlinespace[0.05cm]
        %%%%%%%%%%%%%%%%%%%%%%%%%%%%%%%%%%%%%%%%%%%%%%%%%%%%%%%%%%%%%%%%%%%%%%%%%%%%%
        UK-9. ``Make systems resilient to outages'' & 
        \cite{BITAG1}, \cite{ENISA1}, \cite{CableLabs1}  & 
        & %Actionable?
        DM, OD & 
        1.1, 1.2b \\ \addlinespace[0.05cm]
        %%%%%%%%%%%%%%%%%%%%%%%%%%%%%%%%%%%%%%%%%%%%%%%%%%%%%%%%%%%%%%%%%%%%%%%%%%%%%
        UK-10. ``Monitor system telemetry data'' & 
        \cite{ENISA1}, \cite{CSA2}, \cite{NYC2} & 
        & %Actionable?
        DM, OD, AD & 
        1.1, 1.2b, 3.1a \\ \addlinespace[0.05cm]
        %%%%%%%%%%%%%%%%%%%%%%%%%%%%%%%%%%%%%%%%%%%%%%%%%%%%%%%%%%%%%%%%%%%%%%%%%%%%%
        UK-11. ``Make it easy for consumers to delete personal data'' & 
        \cite{IOTSF2}, \cite{ENISA2},  \cite{OTA1} & 
        \hfil & %Actionable?
        DM, AD & 
        1.1, 1.2b \\ \addlinespace[0.05cm]
        %%%%%%%%%%%%%%%%%%%%%%%%%%%%%%%%%%%%%%%%%%%%%%%%%%%%%%%%%%%%%%%%%%%%%%%%%%%%%
        UK-12. ``Make installation and maintenance of devices easy'' & 
        \cite{ACSCSS1}, \cite{IIC1}, \cite{IOTSF2} & 
        & %Actionable?
        DM, AD & 
        1.1, 1.2a, 1.2b \\ \addlinespace[0.05cm]
        %%%%%%%%%%%%%%%%%%%%%%%%%%%%%%%%%%%%%%%%%%%%%%%%%%%%%%%%%%%%%%%%%%%%%%%%%%%%%
        UK-13. ``Validate input data'' & 
        \cite{ENISA1}, \cite{IOTSF2}, \cite{OWASP2} & 
        \hfil & %Actionable?
        DM, OD, AD & 
        1.2b \\ \addlinespace[0.05cm]
        %%%%%%%%%%%%%%%%%%%%%%%%%%%%%%%%%%%%%%%%%%%%%%%%%%%%%%%%%%%%%%%%%%%%%%%%%%%%%
        \bottomrule
    \end{tabular}
\end{table*}%Practices table

\begin{figure}[t!]
    \centering
    \includegraphics[width=\linewidth]{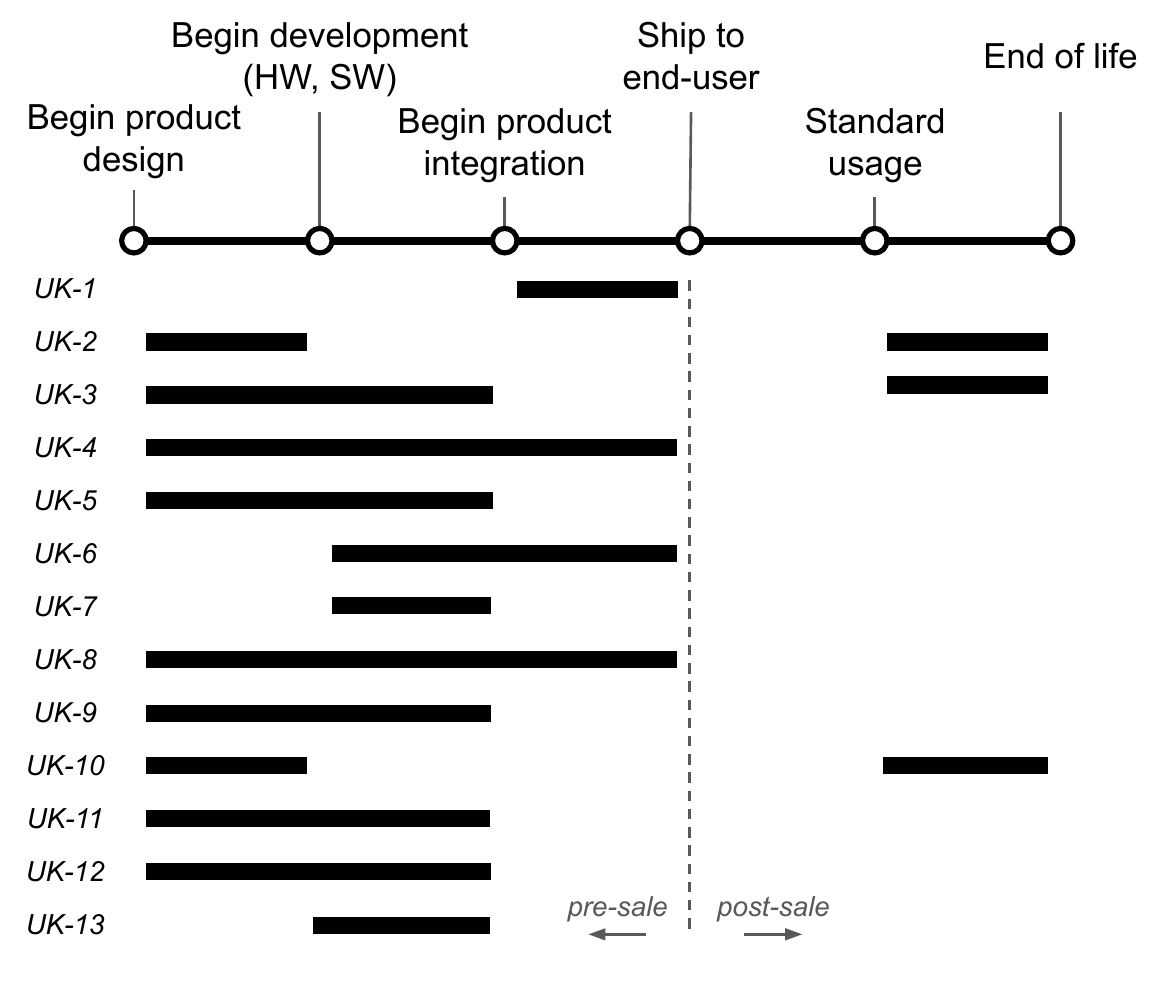}
    \caption{Mapping of DCMS guidelines to selected IoT device lifecycle events.}
    \label{fig:timeline2}
\end{figure}

Table~\ref{tab:practices} includes each DCMS guideline, with a brief listing of example industry documents that incorporates related practices, and where they belong in our four-quadrant grid. We have selected this set of guidelines to highlight their relationship to the IoT lifecycle and how unclear the guidelines are. The DCMS highlights four stakeholders and states that these are ``primarily responsible for implementation''\cite{DCMS2}. Our analysis incorporates a different (but similar and more specific) set of stakeholders (discussed in Section~\ref{sec:stakeholders}), as we feel they more closely fit our model of the IoT lifecycle, particularly with the pre-sale phases where we note a heavy representation of practices (Section~\ref{sec:categorization}). As we note that each of these 13 guidelines are outcomes rather than practices, we classify each of the DCMS' 13 guidelines as either type-V or type-S based on whether or not they imply an action (type-S outcomes suggest actions---Section~\ref{sec:background/actionable}). While some of the 13 are \textit{more} intuitive about what to implement (e.g., \textit{UK-1}), a number of them are rather vague, requiring searching for more depth; or unclear how the problems are solved from a research perspective (e.g., \textit{UK-13}'s secure software update). This relates directly to the concept of outcome versus action, as discussed previously in Section~\ref{sec:background/actionvsoutcome}.

\begin{table*}[t]
    \centering
    \caption{UK Government guidelines \cite{DCMS1} reinterpreted to more clearly suggest actions (i.e., converted \hspace{\textwidth}from type-V outcomes to type-S outcomes). First column maps guidelines to Fig.~\ref{fig:grid} quadrants.}
    \label{tab:guidelineupdates}
    \begin{tabular}{p{0.5cm}p{7.0cm}p{9.0cm}}
        \toprule
        Quad. & DCMS Guideline & Reinterpreted as a Type-S Outcome\\ \midrule
        %%%%%%%%%%%%%%%%%%%%%%%%%%%%%%%%%%%%%%%%%%%%%%%%%%%%%%%%%%%%%%%%%%%%%%%%%%%%%
        1&
        UK-1. ``No default passwords''&
        S-1. If passwords are used, pre-configure with a per-device unique password, rather than a default or no password.
        \\ \addlinespace[0.05cm]
        %%%%%%%%%%%%%%%%%%%%%%%%%%%%%%%%%%%%%%%%%%%%%%%%%
        &
        UK-4. ``Securely store credentials and security-sensitive data''&
        S-4. Ensure user credentials, keys, and other sensitive data is securely stored within the device, through secure hardware storage or modern encryption techniques.
        \\ \addlinespace[0.05cm]
        %%%%%%%%%%%%%%%%%%%%%%%%%%%%%%%%%%%%%%%%%%%%%%%%%
        &
        UK-5. ``Communicate securely''&
        S-5. Maintain confidentiality, integrity, and authentication for communications to/from devices and services.
        \\ \addlinespace[0.05cm]
        %%%%%%%%%%%%%%%%%%%%%%%%%%%%%%%%%%%%%%%%%%%%%%%%%
        &
        UK-6. ``Minimise exposed attack surfaces''&
        S-6. Operate on the principle of least-privilege and disable unused functions such that a device is only capable of running intended functionality.
        \\ \addlinespace[0.05cm]
        %%%%%%%%%%%%%%%%%%%%%%%%%%%%%%%%%%%%%%%%%%%%%%%%%
        &
        UK-8. ``Ensure that personal data is protected'&
        S-8. Use encryption, integrity, and reliable access control mechanisms to protect or back up user data, and comply with data protection laws and regulations for market regions.
        \\ \addlinespace[0.05cm]\midrule
        %%%%%%%%%%%%%%%%%%%%%%%%%%%%%%%%%%%%%%%%%%%%%%%%%%%%%%%%%%%%%%%%%%%%%%%%%%%%%
        %%%%%%%%%%%%%%%%%%%%%%%%%%%%%%%%%%%%%%%%%%%%%%%%%
        2&
        UK-2. ``Implement a vulnerability disclosure policy''&
        S-2. Develop policies specifying actions to follow in the event of a vulnerability being found in your product.
        \\ \addlinespace[0.05cm]
        %%%%%%%%%%%%%%%%%%%%%%%%%%%%%%%%%%%%%%%%%%%%%%%%%
        &
        UK-12. ``Make installation and maintenance of devices easy''&
        S-12. Design devices (physical design and user interfaces) to be conveniently usable for users to facilitate successful installation and maintenance.
        \\ \addlinespace[0.05cm]\midrule
        %%%%%%%%%%%%%%%%%%%%%%%%%%%%%%%%%%%%%%%%%%%%%%%%%%%%%%%%%%%%%%%%%%%%%%%%%%%%%
        %%%%%%%%%%%%%%%%%%%%%%%%%%%%%%%%%%%%%%%%%%%%%%%%%
        3&
        UK-9. ``Make systems resilient to outages''&
        S-9. Form a design, implementation, and test plan for gracefully handling power outages without resulting security exposures.
        \\ \addlinespace[0.05cm]\midrule
        %%%%%%%%%%%%%%%%%%%%%%%%%%%%%%%%%%%%%%%%%%%%%%%%%%%%%%%%%%%%%%%%%%%%%%%%%%%%%
        %%%%%%%%%%%%%%%%%%%%%%%%%%%%%%%%%%%%%%%%%%%%%%%%%
        4&
        UK-3. ``Keep software updated''&
        S-3. Automate and secure the supply and installation of security updates for software/firmware.
        \\ \addlinespace[0.05cm]
        %%%%%%%%%%%%%%%%%%%%%%%%%%%%%%%%%%%%%%%%%%%%%%%%%
        &
        UK-7. ``Ensure software integrity''&
        S-7. Provide means to detect and possibly recover from potential software integrity violations, and return to a known safe-state of the device.
        \\ \addlinespace[0.05cm]
        %%%%%%%%%%%%%%%%%%%%%%%%%%%%%%%%%%%%%%%%%%%%%%%%%
        &
        UK-10. ``Monitor system telemetry data''&
        S-10. Monitor any telemetry or logging data produced by the device for unexpected behavior.
        \\ \addlinespace[0.05cm]
        %%%%%%%%%%%%%%%%%%%%%%%%%%%%%%%%%%%%%%%%%%%%%%%%%
        &
        UK-11. ``Make it easy for consumers to delete personal data''&
        S-11. Provide users with the ability to conveniently view and delete any personal data stored by devices and/or related services.
        \\ \addlinespace[0.05cm]
        %%%%%%%%%%%%%%%%%%%%%%%%%%%%%%%%%%%%%%%%%%%%%%%%%
        &
        UK-13. ``Validate input data''&
        S-13. Ensure safe handling and validation of any input data to prevent malicious inputs, whether by any user, programmatic, or network interface.
        \\ \addlinespace[0.05cm]
        %%%%%%%%%%%%%%%%%%%%%%%%%%%%%%%%%%%%%%%%%%%%%%%%%%%%%%%%%%%%%%%%%%%%%%%%%%%%%
        \bottomrule
    \end{tabular}
\end{table*}

Reinterpreting the previous outcomes, in Table~\ref{tab:guidelineupdates} we produce a matching 13 outcomes that suggest actions, or indicate \textit{more} clearly what steps need to be taken. We consider this a mid-way point between an outcome (specifically a type-V outcome) and an action. As these new outcomes ideally direct the implementer towards actions, we place each into one of the four quadrants from Figure~\ref{fig:grid} and discuss their placement here. Placement of these outcomes are in some cases subjective, but provide an understanding of how the grid relates back to practices. The outcomes discussed in the following four sections (quadrants 1 through 4) discuss these reinterpreted outcomes in the order that they are found in Table~\ref{tab:guidelineupdates}.
    
    %%%%%%%%%%%%%%%%%%%%%%%%%%%%%%%%%%%%%%%%%%%%%%%%%%%%%%%%%%%%%
    \subsection*{Quadrant 1---Easy Implementation, High Effectiveness: Highly Effective Government Regulation/Should Already Be Implemented}
    \label{sec:implement/tr}
    
    Practices that are fairly easy to implement and having high effectiveness make strong candidates for government regulation, but should also be expected to already be implemented. These are the practices that, technically-speaking, are not difficult to implement and have manageable overhead (e.g., low maintenance, high usability, minimal extra services to run), but provide significant improvements to the security of a device. While it should be expected that any manufacturer looking to provide a solid security posture for their products have already implemented these, government regulation may be what is needed to push other non-implementing manufacturers into line. 
    
    \noindent
    \fbox{%
        \begin{minipage}{0.96\linewidth}
            \textit{S-1. If passwords are used, pre-configure with a per-device unique password, rather than a default or no password.}
        \end{minipage}
    }
    
    This outcome seeks to prevent attackers from using a broad scanning technique to try ``default'' (including empty) passwords on wide classes of devices in the hopes that a user has not configured their device, or configured it incorrectly. This was the method of attack that the Mirai botnet \cite{Kolias2017} used to gain access to so many devices. A unique password can be long enough to make it infeasible for attackers to attempt to guess, but short enough to reduce user difficulties in entering a long password. NIST recommends as little as 6 digits for a randomly-chosen PIN for defending against online attacks, assuming other preventative methods are in place (e.g., rate-limiting, failed attempt lock-out)\cite{NISTsp80063}. A prime example of this is the requirement by the California (USA) state government to have non-unique passwords on new IoT devices \cite{Bill327}---the requirement is fairly easy to implement, does not require much maintenance, is easy for the end-users to manage, and adequately defends against default-password-based attacks. Post-sale stakeholders---primarily the end-users---are at risk of having their devices compromised if they are using publicly-known default passwords. Enforcing mandatory unique passwords may incur a small usability penalty if the password is lengthy or complex, but once the first-time configuration and mandatory password change has taken place, the random default password would no longer be used.
    
    \noindent
    \fbox{%
        \begin{minipage}{0.96\linewidth}
            \textit{S-4. Ensure user credentials, keys, and other sensitive data is securely stored within the device, through secure hardware storage or modern encryption techniques.}
        \end{minipage}
    }
    
    A device must be able to protect the data that it is storing. This may include user credentials such as passwords, PINs, and login information; cryptographic keys used for establishing secure communication with services or other devices; or sensitive user-provided/recorded data. In the event that a device is remotely compromised (i.e., an attacker has gained control of the device over a network connection), the attacker must be prevented from accessing this data. In the case of a physical breach where an attacker has physical access to a device, the same applies. Should a device be compromised (whether physically or remotely), an end-user's sensitive data may be leaked, or technical information may be disclosed, allowing an attacker to maliciously use the device. 
    
    \noindent
    \fbox{%
        \begin{minipage}{0.96\linewidth}
            \textit{S-5. Maintain confidentiality, integrity, and authentication for communications to/from devices and services.}
        \end{minipage}
    }
    
    Ensuring communications are secured (i.e., providing end-to-end confidentiality, integrity, and entity authentication) should be expected to be implemented by a manufacturer. Government regulation to enforce such a practice would be useful to ensure manufacturers that are not following this practice. While it may not be the case for many IoT devices, depending on what is being manufactured and the application, some resource constrained devices may struggle to perform certain cryptographic operations useful in secure communications \cite{Bertino2016}. Computation times for public-key operations on low-end 8- or 16-bit processors may be long, but more significantly, the power consumption on battery-powered devices needs to be taken into account.
    
    \noindent
    \fbox{%
        \begin{minipage}{0.96\linewidth}
            \textit{S-6. Operate on the principle of least-privilege and disable unused function such that a device is only capable of running intended functionality.}
        \end{minipage}
    }
    
    It should be an expectation that devices---particularly those that are publicly-accessible or internet-facing---have all non-essential functionality disabled. Assuming this is done adequately, this prevents a device from being used for non-intended purposes such as using non-required protocols (e.g., sending/receiving data via CURL, TELNET to other hosts) or installing new software for additional functionality. This may be difficult to do, however, as it requires complete knowledge of a device and the software running on it. Using a pre-built OS (e.g., some Linux distribution) may require a deeper analysis to find all non-essential functionality and a way to disable it. User accounts in an OS may be one way to short-cut this issue, but any privilege escalation exploits an attacker may be able to use then leaves remaining features free to be used. Wireless sensor network (WSN) or IoT-specific OSs (e.g., RIOT \cite{Baccelli2018}, Contiki \cite{Dunkels2004}) may assist in this with their smaller footprint and specific focus.
    
    \noindent
    \fbox{%
        \begin{minipage}{0.96\linewidth}
            \textit{S-8. Use encryption, integrity, and reliable access control mechanisms to protect or back up user data, and comply with data protection laws and regulations for market regions.}
        \end{minipage}
    }
    
    Personal user data must be protected from being exposed to malicious attackers and non-authorized users. This not only has to do with data stored on the device, but also data in transit and stored on servers, meaning it shares many of the same concerns and issues that B4 and B5 have. Where it differs is that this is information sensitive to the individual users of a product. Data in transit follows the same security requirements as B5 (i.e., communications must maintain confidentiality, integrity, and must only be accessed by authenticated, authorized parties), and stored data (either on a server, hub device, or the generating device itself) must be able to secure data with the same requirements as B4.
    
    %%%%%%%%%%%%%%%%%%%%%%%%%%%%%%%%%%%%%%%%%%%%%%%%%%%%%%%%%%%%%
    \subsection*{Quadrant 2---Easy Implementation, Low Effectiveness: Should be Implemented, But Lower Priority}
    \label{sec:implement/tl}
    
    Practices that are easy to implement but have low effectiveness in preventing vulnerabilities are candidates for what some may consider ``optional'' practices. While best practices for security should generally not be considered optional, we recognize that, particularly from a business perspective, the cost of implementing even an easy practice may still outweigh the benefit, especially if any additional overhead is required to maintain the practice.
    
    \noindent
    \fbox{%
        \begin{minipage}{0.96\linewidth}
            \textit{S-2. Develop policies specifying actions to follow in the event of a vulnerability being found in your product.}
        \end{minipage}
    }
    
    While malicious actors are unlikely to report vulnerabilities in your products, other ethical entities may provide insight into any problem areas that are found, and this should be encouraged. Being notified of existing vulnerabilities makes you both aware of problems and provides an incentive to work to correct the errors, as they are now (as often is the case in academia) publicly-known and potentially exploitable. Making a contact form or address available on a public space (e.g., company website) for users and researchers to report issues is a fairly low-difficulty action to implement, but itself does not necessarily provide highly improved levels of security. Responding and addressing the vulnerabilities is the critical step to be taken.
    
    \noindent
    \fbox{%
        \begin{minipage}{0.96\linewidth}
            \textit{S-12. Design devices (physical design and user interfaces) to be conveniently usable for users to facilitate successful installation and maintenance.}
        \end{minipage}
    }
    
    One of the key characteristics of IoT is their lack of ``standard'' user interfaces, i.e., devices commonly lack screens, keyboards, mice, and touchscreens, instead often relying on remote techniques to interface with devices for purposes other than initial configuration or maintenance. These remote techniques include connecting to a web server being run on the device from an IoC device, or using a manufacturer-provided smartphone/tablet (mobile) app that talks with the IoT device. While this may be somewhat straight-forward given the experience industry and academic researchers have gained over the years by working on mobile and web interfaces, installation and maintenance may now extend to a significant number of devices, each needing their own attention. If devices are not easy to be installed and maintained, users may choose to forgo configuration out of frustration and instead run (if possible) with default configuration. While not directly related, this issue highlights the importance of the ``safe defaults'' design principle.
    
    %%%%%%%%%%%%%%%%%%%%%%%%%%%%%%%%%%%%%%%%%%%%%%%%%%%%%%%%%%%%%
    \subsection*{Quadrant 3---Hard Implementation, Low Effectiveness: Low Return on Investment}
    \label{sec:implement/bl}
    
    Practices that are hard to implement and have low security effectiveness are generally not seen as practices that provide a great return on implementation investment. These practices tend to have a higher level of maintenance (possibly requiring some live service or system to be maintained), a lower usability for the end-users or maintainers, and do not provide significant security benefits to be worth the cost. This ties heavily back into the cost-benefit analysis that an organization would take when investigating practices and their worthiness within IoT products.
    
    \noindent
    \fbox{%
        \begin{minipage}{0.96\linewidth}
            \textit{S-9. Form a design, implementation, and test plan for gracefully handling power outages without resulting security exposures.}
        \end{minipage}
    }
    
    The ability for a device to attempt to prevent service outages is a useful feature, but given how interconnected today's devices are with other local devices or hosts on the internet, resources may be best spent elsewhere. Further, with downtime being quite infrequent in modern services, impact to users---while potentially being critical in the moment for some applications---is likely to not be too severe over a short amount of time. This said, depending on the architecture some devices within an environment may only need basic local transmission to another, perhaps hub device. Ensuring a base level of functionality may be impossible for some devices, but may be remedied by taking actions to reduce the impact when service resumes. For example, if communications with a base station are interrupted, a smart thermometer could maintain readings as normal, but cache each reading until it can regain connectivity with the base station, then forward the data along. This is against the spirit of live and immediate feedback as is common in IoT, but may allow for some functionality such as historical readings to be maintained. Implementations should take care to regain service gracefully, as having to send a large back-log of data may overwhelm receiving devices and potentially drain a significant amount of battery power.

    %%%%%%%%%%%%%%%%%%%%%%%%%%%%%%%%%%%%%%%%%%%%%%%%%%%%%%%%%%%%%
    \subsection*{Quadrant 4---Hard Implementation, High Effectiveness: Prime Target for Research}
    \label{sec:implement/br}
    
    Practices that are difficult to implement, but have high effectiveness, are prime targets for researchers. A practice that has high effectiveness should be a goal to have implemented. It is the role of researchers to find new methods for more easily implementing these currently difficult practices. Ideally, practices in this category would, over time and due to the efforts by researchers, move to the sweet-spot of Easy-High where they could then be regulated (if need-be).
    
    \noindent
    \fbox{%
        \begin{minipage}{0.96\linewidth}
            \textit{S-3. Automate and secure the supply and installation of security updates for software/firmware.}
        \end{minipage}
    }
    
    While effective at eliminating vulnerabilities in devices after being discovered, the implementation of automatic and secure device updates is significantly challenging. If we make the na\"ive assumption that end-users will keep their devices' software updated, this may also be a practice implemented in the Standard Usage phase; however, this is unrealistic. Automatic updates provide pre-sale stakeholders with a mechanism for ensuring that software vulnerabilities can be addressed. While generally not an issue for the IoC, this is a challenge in the IoT for a few reasons. For one, updates must be transmitted securely to ensure the integrity of the update. Highly constrained devices may struggle with public-key cryptography \cite{Bertino2016}, making it more challenging to establish session keys and verify signatures provided with update images. Second, an update infrastructure must be maintained for the lifespan of a product, or at least until a sunset date where a device will be known to not receive any more updates (a separate challenge for IoT). Devices must, in some trusted way, be able to contact update providers for secure transmission. If a manufacturer goes out of business (unlikely for large corporations) or a software provider stops updating their software, reliant devices will now remain un-patched, with any existing vulnerabilities persisting for the life of the device.
    
    \noindent
    \fbox{%
        \begin{minipage}{0.96\linewidth}
            \textit{S-7. Provide means to detect and possibly recover from potential software integrity violations, and return to a known safe-state of the device.}
        \end{minipage}
    }
    
    Devices need to be able to return to some state in which they know is safe. In the case of malicious attacks, implementing defenses to prevent this could be done in hardware \cite{Clercq2016} or by building preventative features into software \cite{Abadi2005}, but consume additional resources wither in the form of slower, more costly computations (particularly on IoT devices where power is a significant resource); or inflating the sizes of binaries and general overhead. In lieu of defenses, the challenge is to know when an integrity violation has taken place. Being able to compare current running processes or files with securely-stored pre-known states may be an avenue, but may not work in cases where pre-known safe states are unknown such as software updates or dynamic content.
    
    \noindent
    \fbox{%
        \begin{minipage}{0.96\linewidth}
            \textit{S-10. Monitor any telemetry or logging data produced by the device for unexpected behavior.}
        \end{minipage}
    }
    
    It is unrealistic to expect end-users to monitor telemetry data produced by their devices, so if information is generated that may be used in preventing or handling attacks, the data must be viewed by the manufacturer (or other party responsible) or done automatically locally. In the former case, the device must be able to report this data to some remote location, requiring services to be kept online for this purpose, and have the ability to securely transmit the data to/from the device. This itself adds another potential attack vector that the manufacturer will have to account for in their threat model. 
    
    Alternatively, intrusion detection systems (IDS) could be employed to protect IoT devices, but this requires either a network-based IDS or one running on each host. Given the constrained nature of IoT devices, this is more challenging; however, recent work has shed light on the potential for IoT-focused IDSs \cite{Mudgerikar2019}\cite{Midi2017}. Efforts are ongoing.
    
    \noindent
    \fbox{%
        \begin{minipage}{0.96\linewidth}
            \textit{S-11. Provide users with the ability to conveniently view and delete any personal data stored by devices and/or related services.}
        \end{minipage}
    }
    
    Ensuring the users can delete any personal data that may have been stored becomes both a usability and technical challenge. For usability, interfaces have to be designed that first allow users to understand the magnitude of the data that may be stored (and potentially \textit{where} it is stored) and to allow them to successfully delete the data. IoT devices typically do not have screens, and those that do tend to not have screens of significant size to allow users to evaluate their data. Interfaces such as voice, small screen, or no interface at all make it difficult to see what data is stored. This is often alleviated by using other devices such as a computer or smartphone to access a website or some other interface in order to review and delete their data, but requires data to be stored in a cloud service, a local hub device, or on a device providing a software interface for external access. 
    
    \noindent
    \fbox{%
        \begin{minipage}{0.96\linewidth}
            \textit{S-13. Ensure safe handling and validation of any input data to prevent malicious inputs, whether by any user, programmatic, or network interface.}
        \end{minipage}
    }
    
    Input data into a process can come from a number of sources including physical user input directly onto the device (e.g., buttons, touch screen, voice commands, etc.), application network data, other pieces of software, or user-facing application programming interfaces (APIs). While it may be somewhat straight-forward to filter out the more obvious inputs such as user text fields (e.g., cross-site scripting, SQL injection attacks), ensuring \textit{all} inputs from other, less obvious sources is more of a challenge. A particularly notable example of this was 2014's Heartbleed vulnerability, which exploited OpenSSL's Heartbeat extension and allowed attackers to retrieve data in memory from beyond the bounds of the message \cite{Durumeric2014}.

%%%%%%%%%%%%%%%%%%%%%%%%%%%%%%%%%%%%%%%%%%%%%%%%%%%%%%%%%%%%%%%%%%%%%%%%%%%%%%%%%%%%%%%%%%%%%%%%%%%%
\section{Related Work}
In this section we highlight related work that highly influenced the topic, discussion, and/or direction of this work. As previously noted, formal definitions for what a ``best practice'' is tend to be lacking in the formal literature. A noted work in this area by King \cite{King2000} was discussed in Section~\ref{sec:bestpractices}. Among papers that define categories of practices or areas of challenges for IoT security, Tschofenig and Baccelli \cite{Tschofenig2019} discuss efforts by ENISA and the IETF to provide recommendations and guidelines for improving IoT security posture. While not defining categories for best practices, that work categorizes technical and organizational areas to be considered for the secure development and usage of IoT devices. Alrawi et al. \cite{Alrawi2019} analyze and systematize work in the field of home-based IoT security and propose a methodology for how the security of home-based systems can be evaluated. Throughout the work they point to related work for reference, but fall short of defining what a best practice is, mentioning that best practices are ``readily available'', but provide no specific references. RFC 8576 \cite{Garcia-Morchon2019_2} proposes a generic model of the lifecycle of an IoT device. Their model does not consider the stakeholders that have a major impact at each phase; however, they acknowledge that it is a simplified model, accurately representing major phases that an IoT device passes through. Looking to how manufacturers can be incentivized to provide a better security posture, Morgner and Benenson \cite{Morgner2018} explore the relationship between efforts in formal IoT technical standards (not to be confused with a ``standard practice'') and the (unfortunate) reality of the economics of IoT security and its implications for the general security posture of manufacturers. 

In the informal literature (not peer-reviewed), despite the lack of best practice definitions, countless documents offer advice. As referenced throughout this work, according to their publicly-provided dataset \cite{IoTSecMap}, the DCMS collected 69 documents from 49 different organizations \cite{DCMS1}. Some of the work more heavily represented in the collection includes security organizations such as the IoT Security Foundation \cite{IoTSF1}, the European Union Agency for Network and Information Security (ENISA) \cite{ENISA1}, the Industrial Internet Consortium (IIC) \cite{IIC1}, and the GSMA \cite{GSMA4}. Industrial security-focused organizations comprise the most highly-referenced sources in the collection; however, a number of government entities (including the previously-mentioned ENISA) and businesses are included. Examples include the US Senate \cite{USSenate1}, the US National Telecommunications and Information Administration (NTIA) \cite{NTIA1}, Microsoft \cite{Microsoft1}, and AT\&T \cite{ATT1}. This is not meant to be an exhaustive list, but represents the breadth of organizations seeking to contribute advice to the IoT security space.

The DCMS targets their 13 guidelines at four stakeholders: device manufacturers, IoT service providers, mobile application developers, and retailers \cite{DCMS2}. Our work focuses on different, more specific stakeholders as they more closely map to the major lifecycle phases presented in Section~\ref{sec:lifecycle}.

%%%%%%%%%%%%%%%%%%%%%%%%%%%%%%%%%%%%%%%%%%%%%%%%%%%%%%%%%%%%%%%%%%%%%%%%%%%%%%%%%%%%%%%%%%%%%%%%%%%%
\section{Discussion \& Concluding Remarks}
The basic concept of best practices is widely known, and understood by many
non-experts even without prior study.  General understanding is an important
foundation from which to build a strong deeper level understanding, but at this
point even the research community lacks a firm basis from which to make
progress.  Indeed as we have argued, in security and technology communities
(not to mention the general public), ambiguity abounds regarding the language
of technical best practices.  We suggest this begins from an absence of clear,
concrete, useful definitions surrounding best practices.  Further compounding
this issue is the wide variety of terms used to describe related concepts.

Our goal has been to improve upon this situation in a systematic way, by
exploring these issues first through generic discussion and analytic
classification, supported and cross-checked through specific focus on consumer
IoT devices and their lifecycle.  We followed this by further analysis of a
large collection of ``security guidelines'' compiled by the UK government.  We
offer a more precise, consistent vocabulary for the community to build on, to
advance the identification of security best practices, and for discussion
within this subject area---one of intense interest to governments and industry,
and of high consequential impact to the general public.

Using our discussion and analysis of a wide selection of practices from
industrial, academic, and government institutions as a viewing lens, we
uncovered conflation of the ideas of security goals (outcomes), and the steps
or methods by which they may be reached (practices). 
We have also noted an important characteristic for a recommendation or guidance:
whether or not it is actionable. Our working definition for ``best practice''
insists that a practice necessarily be actionable.

Reviewing 1014 items of advice (a mix of guidelines, recommendations, standards
and practices) from industrial, governmental, and academic sources
(Section~\ref{sec:categorization}), we find it alarming that 91\% are not
explicit practices that can be followed, but desired outcomes.  We believe that
it is crucially important that the organizations proposing and endorsing these
lists have a clear idea of whether they are recommending practices, or
specifying what might be called baseline security requirements, or simply
offering advice about good principles to think about in the shower.  If the
goal is that relevant stakeholders adopt and implement stated practices towards
the goal of reducing security exposures, then it would appear imperative that
(actionable) best practices be identified and clearly stated, rather than vague
outcomes---lest the target stakeholders find that they are unable to map advice
to a concrete practice, even if they are motivated to do so.

We have also found that the majority (69.6\%) of 1014 recommendations
considered for consumer-focused IoT security are those that would need to be
implemented in the creation (pre-sales) stage of the lifecycle of an IoT
device, leaving these in the hands of the product manufacturer and/or its
hardware and software partners.  End-users, upon purchasing an IoT device, may
positively impact security by following appropriate best practices thereafter,
but are not generally in the position to address deficits created by the
manufacturer. This shines a light on the responsibility of pre-sale
stakeholders, including the overall nameplated manufacturer, to ensure a strong
security posture for items that only they are in a position to control.  Our
analysis thus adds weight to the importance of security motivation for
manufacturers, on whom many of these security practices are incumbent,
but also the importance of identifying, and clearly and
unambiguously stating, (actionable) best practices. Unfortunately, our analysis
suggests that the organizations providing recommendations may well be falling
short of their own goals, if their belief is that their recommendations are
sufficiently detailed to be considered actionable. The main point is: if security experts
do not find that guidelines are easily actionable, then it is unrealistic to
expect that (security non-expert) manufacturers will magically find a way to
adopt and implement the practices.

In order to identify, list and explain instances of best practices, we first
need to agree upon the granularity at which these should be expressed, and
indeed, agree on a useful definition of ``best practice''. Our work shows that
this itself appears to still be a work-in-progress.
Once suitable best practices are identified and captured, a next step is to
take measures that they be adopted and implemented---raising the question of
motivation for individual stakeholders to invest in this exercise.  As
discussed (above and in Section~\ref{sec:categorization}), the pre-sale
stakeholders including manufacturer are essential to success.  A combination of
the economic motivation of manufacturers,
their poor track record in IoT security to date, and their apparent lack of
accountability for security vulnerabilities in general, point to a worrisome future,
keeping in mind also the lessons of markets for lemons \cite{Akerlof1970}.
It seems quite apparent that self-regulation of the IoT industry been largely
unsuccessful; this falls against the backdrop of a grand success of the overall
software industry in disclaiming all liability for software product flaws,
despite itself falling far short of delivering products without security
vulnerabilities.  Whether government regulation of some form will arise, or be
necessary, is a separate question.

The problem of how to arrange that security best practices are adopted and
implemented, across a broad spectrum of hardware-software systems, is not new.
However, the Internet as we know it has moved from an Internet of Computers to
an Internet of Things, with new consequences for the physical world outside of
computers themselves.  IoT promises convenience, productivity, and improved
quality of life, but its new links from the virtual world to our physical
environment and personal safety bring new threats, exacerbated by the promised
pervasiveness of IoT devices. To support security in an IoT world, we believe a
foundational piece is best practices, beginning with a better understanding of
what they are---both generically, and specifically.  One necessary component is
a proper, precise vocabulary that differentiates between desired outcomes, and
methods or practices that help achieve them.  Whether or not manufacturers have
sufficient technical expertise and knowledge, or are sufficiently motivated to
commit resources, to improve IoT security, the research community can play an
important role through first steps in providing the language, and a
foundational understanding, surrounding the meaning, if not instances of,
security best practices.

\textbf{Acknowledgments.}
We thank the members of the Carleton Computer Security Lab, Carleton Internet Security Lab, and anonymous referees for the feedback on this work. Van Oorschot is Canada Research Chair in Authentication and Computer Security, and acknowledges NSERC for funding the chair and a Discovery Grant.

%%%%%%%%%%%%%%%%%%%%%%%%%%%%%%%%%%%%%%%%%%%%%%%%%%%%%%%%%%%%%%%%%%%%%%%%%%%%%%%%%%%%%%%%%%%%%%%%%%%%
\balance
\bibliographystyle{IEEEtran}
\bibliography{bib}

\begin{thebibliography}{10}
\providecommand{\url}[1]{#1}
\csname url@samestyle\endcsname
\providecommand{\newblock}{\relax}
\providecommand{\bibinfo}[2]{#2}
\providecommand{\BIBentrySTDinterwordspacing}{\spaceskip=0pt\relax}
\providecommand{\BIBentryALTinterwordstretchfactor}{4}
\providecommand{\BIBentryALTinterwordspacing}{\spaceskip=\fontdimen2\font plus
\BIBentryALTinterwordstretchfactor\fontdimen3\font minus
  \fontdimen4\font\relax}
\providecommand{\BIBforeignlanguage}[2]{{%
\expandafter\ifx\csname l@#1\endcsname\relax
\typeout{** WARNING: IEEEtran.bst: No hyphenation pattern has been}%
\typeout{** loaded for the language `#1'. Using the pattern for}%
\typeout{** the default language instead.}%
\else
\language=\csname l@#1\endcsname
\fi
#2}}
\providecommand{\BIBdecl}{\relax}
\BIBdecl

\bibitem{Wortmann2015}
F.~Wortmann and K.~Fl{\"u}chter, ``{Internet of Things},'' \emph{Business \&
  Information Systems Engineering}, vol.~57, pp. 221--224, 2015.

\bibitem{Alrawi2019}
O.~Alrawi, C.~Lever, M.~Antonakakis, and F.~Monrose, ``{SoK: Security
  Evaluation of Home-Based IoT Deployments},'' in \emph{{IEEE Symp.\ Security
  and Privacy}}, 2019.

\bibitem{Kolias2017}
C.~Kolias, G.~Kambourakis, A.~Stavrou, and J.~Voas, ``{DDoS in the IoT: Mirai
  and Other Botnets},'' \emph{Computer}, vol.~50, no.~7, pp. 80--84, 2017.

\bibitem{huang2015}
W.~Huang, A.~Ganjali, B.~H. Kim, S.~Oh, and D.~Lie, ``{The State of Public
  Infrastructure-as-a-Service Cloud Security},'' \emph{ACM Computing Surveys
  (CSUR)}, vol.~47, no.~4, pp. 1--31, 2015.

\bibitem{CBC2020}
\BIBentryALTinterwordspacing
E.~Johnson, ``{Online Banking Agreements Protect Banks, Hold Customers Liable
  for Losses, Expert Says},'' 2020. [Online]. Available:
  \url{https://www.cbc.ca/news/business/online-banking-agreements-1.5453192}
\BIBentrySTDinterwordspacing

\bibitem{DCMS1}
\BIBentryALTinterwordspacing
{Department for Digital, Culture, Media \& Sport (DCMS)}, ``{Mapping of IoT
  Security Recommendations, Guidance and Standards to the UK's Code of Practice
  for Consumer IoT Security},'' 2018. [Online]. Available:
  \url{https://assets.publishing.service.gov.uk/government/uploads/system/uploads/attachment_data/file/774438/Mapping_of_IoT__Security_Recommendations_Guidance_and_Standards_to_CoP_Oct_2018.pdf}
\BIBentrySTDinterwordspacing

\bibitem{DCMS2}
\BIBentryALTinterwordspacing
------, ``{Code of Practice for Consumer IoT Security},'' 2018. [Online].
  Available:
  \url{https://assets.publishing.service.gov.uk/government/uploads/system/uploads/attachment_data/file/773867/Code_of_Practice_for_Consumer_IoT_Security_October_2018.pdf}
\BIBentrySTDinterwordspacing

\bibitem{Kruchten2012}
P.~Kruchten, R.~L. Nord, and I.~Ozkaya, ``{Technical Debt: From Metaphor to
  Theory and Practice},'' \emph{IEEE Software}, vol.~29, pp. 18--21, 2012.

\bibitem{rfc1818}
J.~Postel, Y.~Rekhter, and T.~Li, ``{Best Current Practices},'' {IETF}, RFC
  1818, August 1995.

\bibitem{King2000}
G.~King, ``{Best Security Practices: An Overview},'' in \emph{National
  Information Systems Security Conference}, 2000.

\bibitem{CANHandWashing}
\BIBentryALTinterwordspacing
{Health Canada}, ``The benefits of hand washing,'' 2014. [Online]. Available:
  \url{https://www.canada.ca/en/health-canada/services/healthy-living/your-health/diseases/benefits-hand-washing.html}
\BIBentrySTDinterwordspacing

\bibitem{Garcia-Morchon2019_2}
O.~Garcia-Morchon, S.~S. Kumar, and M.~Sethi, ``{RFC8576: Internet of Things
  (IoT) Security: State of the Art and Challenges},'' 2019.

\bibitem{Costin2014}
A.~Costin, J.~Zaddach, A.~Francillon, and D.~Balzarotti, ``{A Large-Scale
  Analysis of the Security of Embedded Firmwares},'' in \emph{USENIX Security
  Symposium}, 2014.

\bibitem{IDBootstrap}
B.~Sarikaya, M.~Sethi, and D.~Garcia-Carrillo, ``{Internet Draft: Secure {IoT}
  Bootstrapping: A Survey},'' 2019.

\bibitem{AUSIoT}
\BIBentryALTinterwordspacing
{Australian Cyber Security Centre}, ``{Securing the Internet of Things for
  Consumers (DRAFT)},'' 2019. [Online]. Available:
  \url{https://www.homeaffairs.gov.au/reports-and-pubs/files/code-of-practice.pdf}
\BIBentrySTDinterwordspacing

\bibitem{rPi4}
\BIBentryALTinterwordspacing
{Raspberry Pi 4}. [Online]. Available:
  \url{https://www.raspberrypi.org/products/raspberry-pi-4-model-b/}
\BIBentrySTDinterwordspacing

\bibitem{arduinoDue}
\BIBentryALTinterwordspacing
{Arduino Due}. [Online]. Available: \url{https://store.arduino.cc/usa/due}
\BIBentrySTDinterwordspacing

\bibitem{rfc7228}
C.~Bormann, M.~Ersue, and A.~Keranen, ``{Terminology for Constrained-Node
  Networks},'' IETF, RFC 7228, May 2014.

\bibitem{Hahm2016}
O.~Hahm, E.~Baccelli, H.~Petersen, and N.~Tsiftes, ``{Operating Systems for
  Low-End Devices in the Internet of Things: A Survey},'' \emph{IEEE Internet
  of Things Journal}, vol.~3, no.~5, pp. 720--734, 2016.

\bibitem{Baccelli2018}
E.~Baccelli, C.~G{\"{u}}ndogan, O.~Hahm, P.~Kietzmann, M.~Lenders, H.~Petersen,
  K.~Schleiser, T.~C. Schmidt, and M.~W{\"{a}}hlisch, ``{RIOT: an open source
  operating system for low-end embedded devices in the IoT},'' \emph{{IEEE}
  Internet of Things Journal}, vol.~5, no.~6, pp. 4428--4440, 2018.

\bibitem{ContikiNG}
\BIBentryALTinterwordspacing
{Contiki-NG: The OS for Next Generation IoT Devices}. [Online]. Available:
  \url{https://github.com/contiki-ng/contiki-ng}
\BIBentrySTDinterwordspacing

\bibitem{Dunkels2004}
A.~{Dunkels}, B.~{Gronvall}, and T.~{Voigt}, ``{Contiki - a Lightweight and
  Flexible Operating System for Tiny Networked Sensors},'' in
  \emph{International Conference on Local Computer Networks}, Nov 2004.

\bibitem{freeRTOS}
\BIBentryALTinterwordspacing
R.~Barry. {The FreeRTOS Kernel}. [Online]. Available:
  \url{https://www.freertos.org/index.html}
\BIBentrySTDinterwordspacing

\bibitem{Ronen2017}
E.~Ronen, A.~Shamir, A.~Weingarten, and C.~O’Flynn, ``{IoT Goes Nuclear:
  Creating a ZigBee Chain Reaction},'' in \emph{IEEE Symposium on Security and
  Privacy}.\hskip 1em plus 0.5em minus 0.4em\relax IEEE, 2017.

\bibitem{Antonakakis2017}
M.~Antonakakis, T.~April, M.~Bailey, E.~Bursztein, J.~Cochran, Z.~Durumeric,
  A.~Halderman, J, D.~Menscher, C.~Seaman, N.~Sullivan, K.~Thomas, and Y.~Zhou,
  ``{Understanding the Mirai botnet},'' in \emph{USENIX Security Symposium},
  2017.

\bibitem{Bill327}
Jackson, ``{United States Senate Bill SB-327 Information Privacy: Connected
  Devices},'' {September 28, 2018}.

\bibitem{GDPR2016}
{The European Parliament and of the Council}, ``{Regulation (EU) 2016/679},''
  {April 27, 2016}.

\bibitem{ENISA1}
\BIBentryALTinterwordspacing
{European Union Agency for Network and Information Security (ENISA)},
  ``{Baseline Security Recommendations for IoT},'' 2017. [Online]. Available:
  \url{https://www.ENISA.europa.eu/publications/baseline-security-recommendations-for-iot}
\BIBentrySTDinterwordspacing

\bibitem{CSA2}
\BIBentryALTinterwordspacing
{Cloud Security Alliance (CSA)}, ``{Security Guidance for Early Adopters of the
  Internet of Things (IoT)},'' 2015. [Online]. Available:
  \url{https://downloads.cloudsecurityalliance.org/whitepapers/Security_Guidance_for_Early_Adopters_of_the_Internet_of_Things.pdf}
\BIBentrySTDinterwordspacing

\bibitem{IEEE1}
\BIBentryALTinterwordspacing
G.~Corser, G.~A. Fink, M.~Aledhari, J.~Bielby, R.~Nighot, S.~Mandal, N.~Aneja,
  C.~Hrivnak, and L.~Cristache, ``{IoT Security Principles and Best
  Practices},'' 2017. [Online]. Available:
  \url{https://internetinitiative.ieee.org/images/files/resources/white_papers/internet_of_things_feb2017.pdf}
\BIBentrySTDinterwordspacing

\bibitem{ACSCSS1}
\BIBentryALTinterwordspacing
G.~Lindsay, B.~Woods, and J.~Corman, ``{Smart Homes and the Internet of
  Things},'' 2016. [Online]. Available:
  \url{http://www.atlanticcouncil.org/images/publications/Smart_Homes_0317_web.pdf}
\BIBentrySTDinterwordspacing

\bibitem{BITAG1}
\BIBentryALTinterwordspacing
{Broadband Internet Technical Advisory Group (BITAG)}, ``{Internet of Things
  (IoT) Security and Privacy Recommendations},'' 2016. [Online]. Available:
  \url{http://www.bitag.org/documents/BITAG_Report_-_Internet_of_Things_(IoT)_Security_and_Privacy_Recommendations.pdf}
\BIBentrySTDinterwordspacing

\bibitem{CableLabs1}
\BIBentryALTinterwordspacing
{CableLabs}, ``{A Vision for Secure IoT},'' 2017. [Online]. Available:
  \url{https://www.cablelabs.com/insights/vision-secure-iot/}
\BIBentrySTDinterwordspacing

\bibitem{AIOTI1}
\BIBentryALTinterwordspacing
{Alliance for Internet of Things Innovation (AIOTI)}, ``{AIOTI Digitisation of
  Industry Policy Recommendations},'' 2016. [Online]. Available:
  \url{https://aioti.eu/wp-content/uploads/2017/03/AIOTI-Digitisation-of-Ind-policy-doc-Nov-2016.pdf}
\BIBentrySTDinterwordspacing

\bibitem{CSA1}
\BIBentryALTinterwordspacing
{Cloud Security Alliance (CSA)}, ``{Future-proofing the Connected World: 13
  Steps to Developing Secure IoT},'' 2016. [Online]. Available:
  \url{https://downloads.cloudsecurityalliance.org/assets/research/internet-of-things/future-proofing-the-connected-world.pdf}
\BIBentrySTDinterwordspacing

\bibitem{AIOTI3}
\BIBentryALTinterwordspacing
{Alliance for Internet of Things Innovation (AIOTI)}, ``{Report on Workshop on
  Security and Privacy in the Hyper connected World},'' 2016. [Online].
  Available:
  \url{https://aioti-space.org/wp-content/uploads/2017/03/AIOTI-Workshop-on-Security-and-Privacy-in-the-Hyper-connected-World-Report-20160616_vFinal.pdf}
\BIBentrySTDinterwordspacing

\bibitem{ATT1}
\BIBentryALTinterwordspacing
{AT\&T}, ``{The CEO's Guide to Securing the Internet of Things},'' 2016.
  [Online]. Available:
  \url{https://www.business.att.com/cybersecurity/docs/exploringiotsecurity.pdf}
\BIBentrySTDinterwordspacing

\bibitem{NYC2}
\BIBentryALTinterwordspacing
{City of New York (NYC) Guidelines for the Internet of Things}, ``{Security},''
  2019. [Online]. Available: \url{https://iot.cityofnewyork.us/security/}
\BIBentrySTDinterwordspacing

\bibitem{IIC1}
\BIBentryALTinterwordspacing
S.~Schrecker, H.~Soroush, J.~Molina, J.~LeBlank, F.~Hirsch, M.~Buchheit,
  A.~Ginter, R.~Martin, H.~Banavara, S.~Eswarahally, K.~Raman, A.~King,
  Q.~Zhang, P.~KacKay, and B.~Witten, ``{Industrial Internet of Things Volume
  G4: Security Framework v1.0},'' 2016. [Online]. Available:
  \url{https://www.iiconsortium.org/pdf/IIC_PUB_G4_V1.00_PB-3.pdf}
\BIBentrySTDinterwordspacing

\bibitem{GSMA4}
\BIBentryALTinterwordspacing
{GSMA}, ``{IoT Security Guidelines for Endpoint Ecosystems---Version 2.0},''
  2017. [Online]. Available:
  \url{https://www.gsma.com/iot/wp-content/uploads/2017/10/CLP.13-v2.0.pdf}
\BIBentrySTDinterwordspacing

\bibitem{AIOTI4}
\BIBentryALTinterwordspacing
{Alliance for Internet of Things Innovation (AIOTI)}, ``{Report: Working Group
  4---Policy},'' 2015. [Online]. Available:
  \url{https://aioti.eu/wp-content/uploads/2017/03/AIOTIWG04Report2015-Policy-Issues.pdf}
\BIBentrySTDinterwordspacing

\bibitem{NYC1}
\BIBentryALTinterwordspacing
{City of New York (NYC) Guidelines for the Internet of Things}, ``{Privacy +
  Transparency},'' 2019. [Online]. Available:
  \url{https://iot.cityofnewyork.us/privacy-and-transparency/}
\BIBentrySTDinterwordspacing

\bibitem{IOTSF2}
\BIBentryALTinterwordspacing
A.~Soorya, A.~Marguilis, A.~Sambordaran, C.~Hills, C.~Shire, G.~Markall,
  I.~Phillips, I.~Dangana, J.~Krueger, J.~Bennett, J.~Moor, L.~Johri,
  M.~Beaumont, N.~Hayes, P.~Gupta, P.~Burgers, R.~Marshall, R.~Storer,
  R.~Dobson, R.~Shepherd, S.~Gulliford, and T.~Hall, ``{IoT Security Compliance
  Framework 2.0},'' 2018. [Online]. Available:
  \url{https://www.iotsecurityfoundation.org/wp-content/uploads/2018/12/IoTSF-IoT-Security-Compliance-Framework-Release-2.0-December-2018.pdf}
\BIBentrySTDinterwordspacing

\bibitem{ENISA2}
\BIBentryALTinterwordspacing
{European Union Agency for Network and Information Security (ENISA)},
  ``{Security and Resilience of Smart Home Environments},'' 2015. [Online].
  Available:
  \url{https://www.ENISA.europa.eu/publications/security-resilience-good-practices}
\BIBentrySTDinterwordspacing

\bibitem{OTA1}
\BIBentryALTinterwordspacing
{Online Trust Alliance (OTA)}, ``{IoT Security \& Privacy Trust Framework
  v2.5},'' 2017. [Online]. Available:
  \url{https://www.internetsociety.org/wp-content/uploads/2018/05/iot_trust_framework2.5a_EN.pdf}
\BIBentrySTDinterwordspacing

\bibitem{OWASP2}
\BIBentryALTinterwordspacing
{Open Web Application Security Project (OWASP)}, ``{OWASP Secure Coding
  Practices Quick Reference Guide},'' 2010. [Online]. Available:
  \url{https://www.owasp.org/images/0/08/OWASP_SCP_Quick_Reference_Guide_v2.pdf}
\BIBentrySTDinterwordspacing

\bibitem{NISTsp80063}
P.~A. Grassi, J.~L. Fenton, E.~M. Newton, R.~A. Perlner, A.~R. Regenscheid,
  W.~E. Burr, J.~P. Richer, N.~B. Lefkovitz, J.~M. Danker, Y.-Y. Choong, K.~K.
  Greene, and M.~F. Theofanos, ``{SP 800-63B---Digital Identity Guidelines:
  Authentication and Lifecycle Management},'' National Institute of Standards
  and Technology (NIST), Tech. Rep., 2017.

\bibitem{Bertino2016}
E.~Bertino, ``{Data Security and Privacy in the IoT},'' in \emph{EDBT}, 2016.

\bibitem{Clercq2016}
R.~de~Clercq, J.~G{\"o}tzfried, D.~{\"U}bler, P.~Maene, and I.~Verbauwhede,
  ``{SOFIA: Software and Control Flow Integrity Architecture},'' \emph{Design,
  Automation \& Test in Europe (DATE)}, pp. 1172--1177, 2016.

\bibitem{Abadi2005}
M.~Abadi, M.~Budiu, {\'U}.~Erlingsson, and J.~Ligatti, ``Control-flow
  integrity,'' in \emph{ACM CCS}, 2005.

\bibitem{Mudgerikar2019}
A.~Mudgerikar, P.~Sharma, and E.~Bertino, ``{E-Spion: A System-Level Intrusion
  Detection System for IoT Devices},'' in \emph{ACM Asia CCS}, 2019.

\bibitem{Midi2017}
D.~Midi, A.~Rullo, A.~Mudgerikar, and E.~Bertino, ``{Kalis—A System for
  Knowledge-Driven Adaptable Intrusion Detection for the Internet of Things},''
  in \emph{International Conference on Distributed Computing Systems (ICDCS)},
  2017.

\bibitem{Durumeric2014}
Z.~Durumeric, J.~Kasten, D.~Adrian, J.~A. Halderman, M.~Bailey, F.~Li,
  N.~Weaver, J.~Amann, J.~G. Beekman, M.~Payer, and V.~Paxson, ``{The Matter of
  Heartbleed},'' in \emph{IMC '14}, 2014.

\bibitem{Tschofenig2019}
H.~Tschofenig and E.~Baccelli, ``{Cyberphysical Security for the Masses: A
  Survey of the Internet Protocol Suite for Internet of Things Security},''
  \emph{IEEE Security {\&} Privacy}, vol.~17, no.~5, pp. 47--57, Sep 2019.

\bibitem{Morgner2018}
P.~Morgner and Z.~Benenson, ``{Exploring Security Economics in IoT
  Standardization Efforts},'' in \emph{Workshop on Decentralized IoT Security
  and Standards (DISS)}, 2018.

\bibitem{IoTSecMap}
\BIBentryALTinterwordspacing
{Copper Horse Ltd.}, ``{Mapping Security \& Privacy in the Internet of
  Things},'' 2019. [Online]. Available: \url{https://iotsecuritymapping.uk/}
\BIBentrySTDinterwordspacing

\bibitem{IoTSF1}
\BIBentryALTinterwordspacing
{IoT Security Foundation}, ``{IoT Security Compliance Framework 1.1},'' 2017.
  [Online]. Available:
  \url{https://www.iotsecurityfoundation.org/wp-content/uploads/2017/12/IoT-Security-Compliance-Framework_WG1_2017.pdf}
\BIBentrySTDinterwordspacing

\bibitem{USSenate1}
\BIBentryALTinterwordspacing
{US Senate}, ``{Bill---S.1691 - Internet of Things (IoT) Cybersecurity
  Improvement Act of 2017 (Bill)},'' 2017. [Online]. Available:
  \url{https://www.congress.gov/bill/115th-congress/senate-bill/1691/text?format=txt}
\BIBentrySTDinterwordspacing

\bibitem{NTIA1}
\BIBentryALTinterwordspacing
{US National Telecommunications and Information Administration (NTIA)},
  ``{Voluntary Framework for Enhancing Update Process Security},'' 2017.
  [Online]. Available:
  \url{https://www.ntia.doc.gov/files/ntia/publications/ntia_iot_capabilities_oct31.pdf}
\BIBentrySTDinterwordspacing

\bibitem{Microsoft1}
\BIBentryALTinterwordspacing
{Microsoft}, ``{Security best practices for Internet of Things (IoT)},'' 2018.
  [Online]. Available:
  \url{https://docs.microsoft.com/en-us/azure/iot-fundamentals/iot-security-best-practices}
\BIBentrySTDinterwordspacing

\bibitem{Akerlof1970}
G.~A. Akerlof, ``{The Market for ``Lemons'': Quality Uncertainty and the Market
  Mechanism},'' \emph{The Quarterly Journal of Economics}, vol.~84, no.~3, pp.
  488--500, 1970.

\end{thebibliography}
%%%%%%%%%%%%%%%%%%%%%%%%%%%%%%%%%%%%%%%%%%%%%%%%%%%%%%%%%%%%%%%%%%%%%%%%%%%%%%%%%%%%%%%%%%%%%%%%%%%%

\end{document}